\documentclass[twocolumn]{jpsj3}
\usepackage{txfonts}
\RequirePackage{graphicx,color}

\title{
Theory of Field-Angle-Resolved Magnetoacoustic Resonance in Spin-Triplet Systems
for Application to Nitrogen-Vacancy Centers in Diamond
}

\author{Mikito Koga$^1$ and Masashige Matsumoto$^2$}

\inst{
$^1$Department of Physics, Faculty of Education, Shizuoka University, Shizuoka
422--8529, Japan \\
$^2$Department of Physics, Faculty of Science, Shizuoka University, Shizuoka
422--8529, Japan \\
}

\abst{ 
Motivated by the recent studies of acoustically driven electron spin resonance applied to diamond
nitrogen-vacancy (NV) centers, we investigate the interaction of an electronic spin-triplet state with
periodically time-dependent oscillating strain fields.
On the basis of a lowest-lying two-level system, we show the importance of two-phonon transition
probabilities controlled by rotating an applied magnetic field using the Floquet theory.
In particular, we demonstrate how to evaluate coupling-strength parameters in the spin--strain
interaction for the $C_{3v}$ point group considering the NV spin states.
The level splitting of spin states can be adjusted by changing the field directions relative to
the NV axis to obtain lower phonon resonance frequencies suitable for practical applications.
Focusing on a field-rotation angle for the vanishment of a longitudinal phonon-mediated transition,
we show that the magnetoacoustic resonance presented here provides useful information as a
new probe of unquantified spin--strain couplings possessed by NV defects.
}

\begin{document}

\maketitle

\newcommand{\ds}{\displaystyle}

\renewcommand{\H}{\mathcal{H}}
\newcommand{\br}{{\mbox{\boldmath$r$}}}
\newcommand{\bR}{{\mbox{\boldmath$R$}}}
\newcommand{\bS}{{\mbox{\boldmath$S$}}}
\newcommand{\bk}{{\mbox{\boldmath$k$}}}
\newcommand{\bH}{{\mbox{\boldmath$H$}}}
\newcommand{\bh}{{\mbox{\boldmath$h$}}}
\newcommand{\bJ}{{\mbox{\boldmath$J$}}}
\newcommand{\bI}{{\mbox{\boldmath$I$}}}
\newcommand{\bPsi}{{\mbox{\boldmath$\Psi$}}}
\newcommand{\bpsi}{{\mbox{\boldmath$\psi$}}}
\newcommand{\bPhi}{{\mbox{\boldmath$\Phi$}}}
\newcommand{\bd}{{\mbox{\boldmath$d$}}}
\newcommand{\bG}{{\mbox{\boldmath$G$}}}
\newcommand{\bu}{{\mbox{\boldmath$u$}}}
\newcommand{\be}{{\mbox{\boldmath$e$}}}
\newcommand{\bE}{{\mbox{\boldmath$E$}}}
\newcommand{\bp}{{\mbox{\boldmath$p$}}}
\newcommand{\bB}{{\mbox{\boldmath$B$}}}
\newcommand{\om}{{\omega_n}}
\newcommand{\omm}{{\omega_{n'}}}
\newcommand{\omd}{{\omega^2_n}}
\newcommand{\omt}{{\tilde{\omega}_{n}}}
\newcommand{\ommt}{{\tilde{\omega}_{n'}}}
\newcommand{\btau}{{\hat{\tau}}}
\newcommand{\brho}{{\mbox{\boldmath$\rho$}}}
\newcommand{\bsigma}{{\mbox{\boldmath$\sigma$}}}
\newcommand{\bSigma}{{\mbox{\boldmath$\Sigma$}}}
\newcommand{\bt}{{\hat{t}}}
\newcommand{\bq}{{\hat{q}}}
\newcommand{\bLambda}{{\hat{\Lambda}}}
\newcommand{\bDelta}{{\hat{\Delta}}}
\newcommand{\bU}{{\hat{U}}}
\newcommand{\bskp}{{\mbox{\scriptsize\boldmath $k$}}}
\newcommand{\skp}{{\mbox{\scriptsize $k$}}}
\newcommand{\bsrp}{{\mbox{\scriptsize\boldmath $r$}}}
\newcommand{\bsRp}{{\mbox{\scriptsize\boldmath $R$}}}
\newcommand{\bsk}{\bskp}
\newcommand{\sk}{\skp}
\newcommand{\bsr}{\bsrp}
\newcommand{\bsR}{\bsRp}
\newcommand{\ri}{{\rm i}}
\newcommand{\re}{{\rm e}}
\newcommand{\rd}{{\rm d}}
\newcommand{\rM}{{\rm M}}
\newcommand{\rs}{{\rm s}}
\newcommand{\rt}{{\rm t}}

\section{Introduction}
It is not unusual that solid-state spins couple to acoustic waves or mechanical oscillators through
the spin--strain interaction when the spin is represented by a spin operator $\bS$ with $S \ge 1$.
\cite{VanVleck40,Donoho64,Barfuss15,Lee17,Udvarhelyi18,Barfuss19,Chen20}
Such spin--strain coupling is analogous to the coupling between electric quadrupoles and
local strains owing to the crystal lattice deformations.
\cite{Mitsumoto14}
In the absence of space inversion symmetry at the spin site, the spin behaves like an electric
dipole and also couples to electric fields.
\cite{Kiel72,Mims76,VanOort90,Doherty12,Matsumoto17}
The additional spin properties concerning the quadrupoles suggest great potential in the acoustic
or electric control of quantum spin states as an alternative method to coherent spin control using
resonant microwaves or light.
\cite{MacQuarrie13,Klimov14,Ovartchaiyapong14}
Optomechanical quantum control can also be realized by coupling electronic spin or orbital states
simultaneously to both laser and acoustically driven strain fields.
\cite{Kepesidis13,Golter16,Chen18}
\par

As a promising platform for various applications to spin control, nitrogen-vacancy (NV) centers in
diamond have a long coherence time that guarantees the robustness of the spin states at room
temperature.
\cite{Balasubramanian09,Mizuochi09,Herbschleb19} 
This is of great advantage to the practical use of quantum information processing and quantum
sensing applications.
\cite{Suter17,Degen17,Rembold20,Ruf21}
For the NV spin, the $S = 1$ triplet ground state is split into the lowest singlet $| S_z = 0 \rangle$
and higher doublet $| S_z = \pm 1 \rangle$ states under a uniaxial crystal field environment with
threefold symmetry.
There are two types of quantum spin transitions.
One is the double quantum (DQ) spin transition ($| +1 \rangle \leftrightarrow | -1 \rangle$),
which has frequently been considered as a primary coupling of the phonon-driven quantum
transition via the quadrupoles.
Rabi oscillations of NV spin states were observed by mechanically controlled measurements
using high-frequency acoustic waves on diamond substrates.
\cite{MacQuarrie15}
The other is the single quantum (SQ) spin transition
($| 0 \rangle \leftrightarrow | \pm 1 \rangle$) usually associated with magnetic control using
resonant microwaves.
In a conventional way, it is necessary to combine magnetically and mechanically driven
quantum transitions to control all the transitions between the sublevels of the NV spin.
Currently, both SQ and DQ spin transitions can be controlled by only magnetically driven photon
fields.
For instance, the DQ transition can be realized by the application of microwave multipulse
sequences with two frequency components.
\cite{Mamin14,Bauch18,Barry20}
Recently, it has been achieved by combined microwave and radio-frequency pulse sequences
under a magnetic field applied orthogonal to the NV axis.
\cite{Yamaguchi20}
\par

Besides such tremendous progress in optical quantum control, it is worthwhile to explore acoustic
quantum control of local spins embedded in a solid.
In particular, we will pursue various possible applications of optomechanical quantum control
using phonon-assisted photon transitions or photon-assisted phonon transitions in a wide range
of frequency.
\cite{Kepesidis13,Golter16,Dong19}
For this purpose, it is highly important to elucidate the phonon-driven SQ spin transition, although
the corresponding spin--strain coupling effect on the NV energy level shift has been considered
to be negligibly small, by analogy with the electric-field coupling to the NV spin.
\cite{VanOort90,Doherty12,Michl19,Kehayias19}
This assumption is valid for static or low-frequency strain fields when the levels $| 0 \rangle$ and
$| \pm 1 \rangle$ are well separated.
However, it is probably not the case for a resonant strain field driven by a high-frequency acoustic
wave of gigahertz order even in the absence of a magnetic field.
For the phonon-driven SQ spin transition, the first challenging experiment to quantify the unknown
coupling strength has been performed by Rabi spectroscopy using uniaxial stress fields.
\cite{Chen20}
It is reported that the corresponding spin--stress coupling strength is not negligibly small
and one order of magnitude larger than a theoretically evaluated value.
\cite{Udvarhelyi18}
More detailed investigations are required to gain a complete understanding of all NV spin
properties driven by strain fields according to a full description of symmetry-allowed spin--strain
interaction in Ref.~5.
\par

In our previous study, we presented a theory of magnetoacoustic resonance (MAR) for ultrasonic
measurements to probe spin--strain couplings considering the NV defect.
\cite{Koga20b}
The important point is that the coupling-strength parameters can be evaluated from the
dependence of the two-phonon transition probability on the rotation angle of an applied magnetic
field around the NV axis.
A two-phonon transition requires longitudinal couplings that do not participate in a single-phonon
transition via transverse couplings.
The longitudinal coupling is represented by a diagonal matrix element for the couplings between
two energy levels, while the transverse coupling is represented by an off-diagonal transition matrix
element.
This is an analogue of a two-photon magnetic transition that requires not only a circularly
polarized field ($\sigma^\pm$ photon) but also a linearly polarized field ($\pi$ photon).
\cite{Gromov00}
Similarly, a single acoustic wave induces various strain-field components, and generates both
longitudinal and transverse couplings.
For phonon transitions, each coupling strength can be controlled by rotating a magnetic field
around the principal axis.
\cite{Koga20b,Matsumoto20,Koga20a}
A two-phonon transition can be detected as well as a single-phonon transition, although the
former has been considered to be more difficult to probe.
Recently, multiphonon transitions have been observed in experiments using an acoustic
resonator coupled to a superconducting qubit.
\cite{Valimaa21}
For the NV center, phonon-driven orbital transitions were detected by photoluminescence
excitation spectroscopy, and the observed phonon sidebands were well explained by
quantum transitions between multiphonon dressed states (see Sect.~3.2).
\cite{Chen18}
For the sublevels of the NV spin, we also expect the possibility of such multiphonon-mediated
optical transitions.
\cite{Kepesidis13,Golter16,Dong19}
Theoretically, a phonon-assisted optical transition can be exchanged for a photon-assisted
acoustically driven transition when the photon frequency is much lower than the phonon frequency.
\cite{Koga20a}
A multiphonon transition also becomes more prominent for a high-frequency acoustically driven
strain field.
\par

In this paper, we focus on the two-phonon transitions considering the spin--strain interaction of the
NV center as the first step toward future MAR measurements, which is an extension of our
previous studies.
\cite{Koga20b,Matsumoto20,Koga20a}
On the basis of the two lowest-lying levels of the NV spin being coupled to periodically
time-dependent oscillating fields, we demonstrate how to evaluate spin--strain coupling-strength
parameters from the two-phonon transition probabilities using the Floquet theory.
\cite{Shirley65,Son09}
We change the direction of the applied magnetic field toward the NV axis to
obtain lower resonant frequencies for ultrasonic measurements.
This is also important for the evaluation of the spin--strain coupling-strength parameters from the
two-phonon transition probabilities.
\par

This paper is organized as follows.
In Sect.~2, we first present a spin--strain interaction Hamiltonian for the $S=1$ electronic states
with $C_{3v}$ symmetry to describe the strain-field couplings using quadrupole (second-rank
tensorial) operators.
The matrix elements of quadrupole--strain coupling are given for the three sublevels of $S=1$
coupled to oscillating strain fields.
In Sect.~3, we formulate the calculation of multiphonon transition probabilities between the two
levels on the basis of a matrix form of the Floquet Hamiltonian.
The evaluation of various spin--strain couplings is demonstrated by investigating the two-phonon
transition probability as a function of the rotation angle of a magnetic field in Sect.~4.
In particular, we focus on an important role of longitudinal couplings in the vanishment of the
two-phonon transition.
This argument is applied to the NV spin state to evaluate the spin--strain coupling-strength
parameters.
The last section gives a summary of this paper and a possible extension of the present study.
In Appendix~A, we show an original form of the spin--strain interaction Hamiltonian for the
$C_{3v}$ point group.
The coupling-strength parameters are transformed between the $C_{3v}$ and cubic crystal
reference frames.
This transformation is applied to the spin--stress interaction as well.
In Appendix~B, we describe the quadrupole--strain couplings under the magnetic field
perpendicular to the NV axis as a simple case.
In Appendix~C, we discuss the applicability of the present analysis to multiple components of
strain fields driven by a surface acoustic wave.
Finally, we show that the higher excitation level in the spin state only exerts a minor effect on the
lowest-lying two-level quantum transitions under a relatively strong magnetic field in Appendix~D.

\section{Model}
\subsection{Spin--strain interaction with $C_{3v}$ symmetry for NV spin}
For the NV electronic spin states in the $C_{3v}$ crystalline-electric-field environment, we study
essential features of quadrupole--strain coupling using the following simplified spin--strain
interaction Hamiltonian:
\cite{Koga20b}
\begin{align}
H_\varepsilon = \sum_k A_{k, \varepsilon} O_k~~(k = u, v, zx),
\label{eqn:Hep}
\end{align}
where the quadrupole operators $O_k$ are constructed by the $S = 1$ spin operator
$\bS = (S_x, S_y, S_z)$ as
\begin{align}
& O_u = \frac{1}{\sqrt{3}} ( 2 S_z^2 - S_x^2 - S_y^2 ) = \frac{1}{\sqrt{3}} [ 3 S_z^2 - S (S + 1) ],
\nonumber \\
& O_v = S_x^2 - S_y^2,~~O_{zx} = S_z S_x + S_x S_z.
\label{eqn:Ok}
\end{align}
For the $C_{3v}$ coordinates,  we choose the NV axis along $\be_z = (1,1,1)/\sqrt{3}$ and
the two orthogonal axes along $\be_y = (1,-1,0)/\sqrt{2}$ and $\be_x = (-1,-1,2)/\sqrt{6}$.
In general, there are five quadrupole components involved in the spin--strain interaction.
Here, we consider only three of them in Eq.~(\ref{eqn:Ok}) for brevity.
Note that $O_v$ and $O_{zx}$ are responsible for the DQ
($| + 1 \rangle \leftrightarrow | - 1 \rangle$) and SQ ($| 0 \rangle \leftrightarrow | \pm 1 \rangle$)
transitions mediated by phonons, respectively.
We express the strain-dependent coupling coefficients $A_{k, \varepsilon}$ in terms of the cubic
crystal ($XYZ$) coordinates ($X \parallel [100]$, $Y \parallel [010]$, and $Z \parallel [001]$) as
\begin{align}
& A_{u, \varepsilon} = g_a \varepsilon_1
~~[ \varepsilon_1 \equiv ( \varepsilon_{YZ} + \varepsilon_{ZX} + \varepsilon_{XY} ) / \sqrt{3} ],
\nonumber \\
& A_{v,\varepsilon} = ( g_b \varepsilon_{U_1} + g_c \varepsilon_{U_2} ) / \sqrt{3},~
A_{zx,\varepsilon} = ( 2 g_d \varepsilon_{U_1} - g_e \varepsilon_{U_2} ) / \sqrt{6},
\nonumber \\
& [ \varepsilon_{U_1} \equiv( 2 \varepsilon_{ZZ} - \varepsilon_{XX} - \varepsilon_{YY} ) / \sqrt{3},~
\varepsilon_{U_2} \equiv ( 2 \varepsilon_{XY} - \varepsilon_{YZ} - \varepsilon_{ZX} ) / \sqrt{3} ],
\label{eqn:Ak-ep}
\end{align}
where the strain tensors
\begin{align}
\varepsilon_{i j} = \frac{1}{2} \left( \frac{\partial u_i}{\partial x_j}
 + \frac{\partial u_j}{\partial x_i} \right)
\end{align}
are related to the displacement vector $\bu = (u_1, u_2, u_3) = (u_X, u_Y, u_Z)$, and
$(x_1, x_2, x_3) = (X,Y,Z)$.
The three symmetry components of oscillating strain fields are shown schematically in
Fig.~\ref{fig:1}(a) on the right.
Indeed, two constraints, $\varepsilon_{XX} =\varepsilon_{YY}$ and
$\varepsilon_{YZ} = \varepsilon_{ZX}$, have been introduced to the lattice deformations because
the three quadrupole components in Eq.~(\ref{eqn:Ok}) are restricted in the spin--strain interaction.
There are five independent coupling parameters $g_i$ ($i = a, b, c, d, e$), and a bulk strain
component has been disregarded.
In Appendix~A, we show a different representation of $A_{k, \varepsilon}$ in terms of the strain
tensors in the $C_{3v}$ reference frame and the relationship between the two representations of
coupling-strength parameters in Eq.~(\ref{eqn:gh}).
\par

\subsection{Quadrupole-strain coupling to NV spin states in a magnetic field}
\begin{figure}
\begin{center}
\includegraphics[width=7cm,clip]{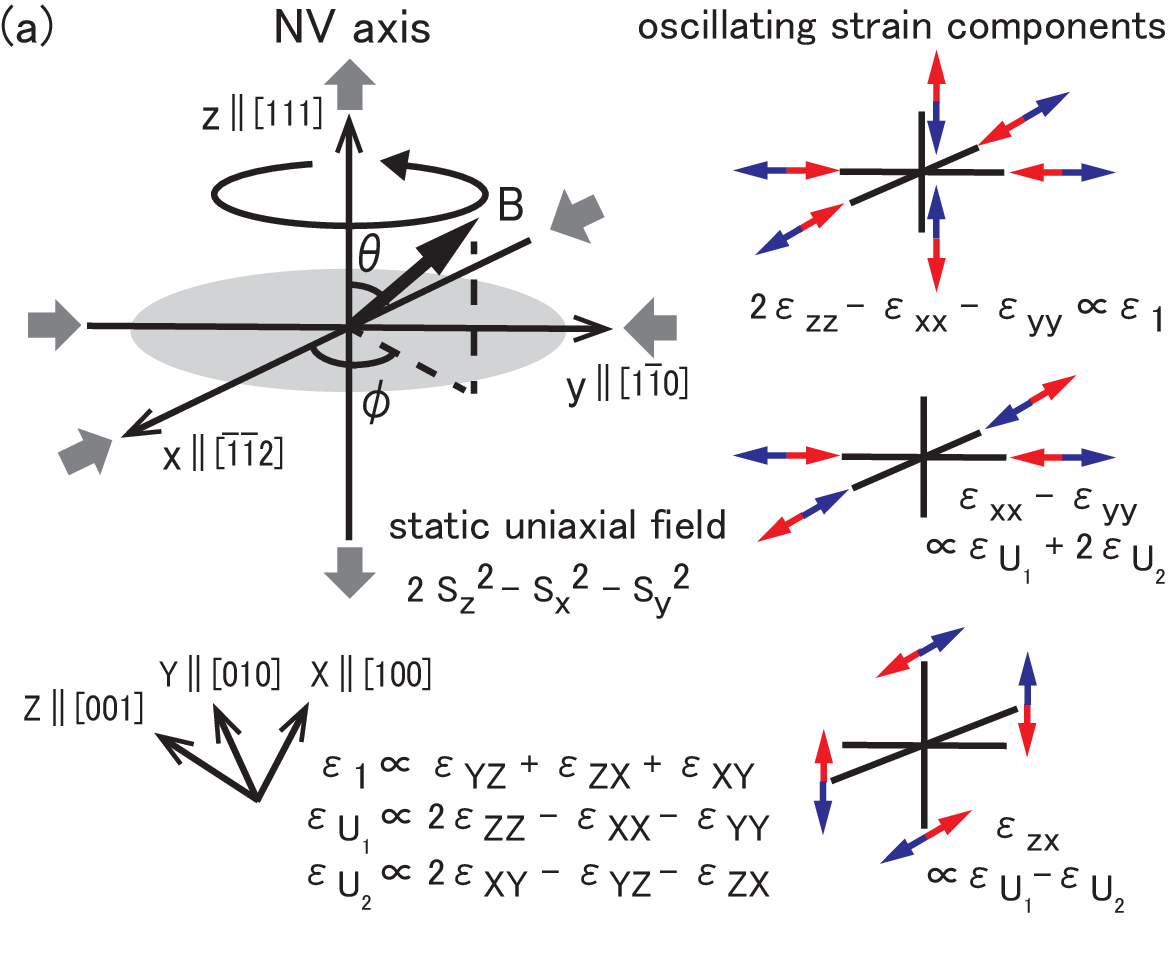}
\includegraphics[width=6cm,clip]{fig1b.eps}
\end{center}
\caption{
(Color online)
(a) Left: Illustration for the rotation of magnetic field $B$ around the NV axis
($z \parallel [111]$), applied to the spin-triplet system.
The polar angle $\theta$ of the field is changed to adjust the level splitting of the electronic spin
states.
The field-rotation angle $\phi$ is measured from the $x$ axis $[ \bar{1} \bar{1} 2 ]$ in the $C_{3v}$
($xyz$) reference frame.
The static uniaxial field is denoted by the thick arrows.
Right: Three symmetry components of oscillating strain fields for MAR considered here.
$\varepsilon_1$, $\varepsilon_{U_1}$, and $\varepsilon_{U_2}$ are the representations in
the cubic ($XYZ$) reference frame.
The double-headed arrows denote the oscillations of each strain component.
(b) Energies $\varepsilon_{12}$ (solid line) and $\varepsilon_{13}$ (broken line) of the
first and second excited states measured from the ground-state energy in the triplet, respectively,
plotted as a function of $\theta$ ($0.08 < \theta / \pi < 0.3$).
Here, $D_0 / \omega = 0.6$, $\gamma_e B / \omega = 2.0$, and the level splitting is normalized
by the frequency $\omega$ of the acoustic wave.
}
\label{fig:1}
\end{figure}
Here, we consider the $S=1$ electronic states in a uniaxial crystal field along the threefold
axis.
The energies and wave functions of the local state depend on the directions of the applied
magnetic field, as shown in Fig.~\ref{fig:1}(a) on the left.
The corresponding Hamiltonian is given by
\begin{align}
H_{\rm l} = - \gamma_e \bB \cdot \bS  + \sqrt{3} D_0 O_u,
\end{align}
where $\bB = B ( \sin \theta \cos \phi, \sin \theta \sin \phi, \cos \theta)$ is an external magnetic field,
$\gamma_e = 2.8$ MHz/G is the electron gyromagnetic ratio, and $D$ ($\equiv 3D_0 > 0$) equals
the energy of the $S=1$ doublet excited state measured from the singlet ground state for $B = 0$.
On the basis of the three states ($S_z = 1, 0, -1$), the eigenvalue problem of $H_{\rm l}$ is
solved using $\tilde{H}_{\rm l} \vec{C}_\mu = \bar{E}_\mu \vec{C}_\mu$, where
\begin{align}
& \tilde{H}_{\rm l} =
\left(
\begin{array}{ccc}
1 - \bar{b} \cos \theta & - ( \bar{b} / \sqrt{2} ) \sin \theta & 0 \\
- ( \bar{b} / \sqrt{2} ) \sin \theta & - 2 &  - ( \bar{b} / \sqrt{2} ) \sin \theta \\
0 & - ( \bar{b} / \sqrt{2} ) \sin \theta & 1 + \bar{b} \cos \theta
\end{array}
\right),
\nonumber \\
& \vec{C}_\mu = ( C_{1 \mu}, C_{0 \mu}, C_{-1, \mu} )^t.
\label{eqn:bHl}
\end{align}
The three eigenstates are labelled as $\mu$ ($= 1, 2, 3$) and
$\bar{E}_1 \le \bar{E}_2 \le \bar{E}_3$ for their energy levels.
$\bar{E}_\mu$ is normalized by $D_0$ and $\bar{b} = \gamma_e B / D_0$.
Using $C_{m \mu}$ ($m = 1, 0, -1)$ for $\bar{E}_\mu$, the wave function is given by
\begin{align}
| \psi_\mu \rangle = C_{1 \mu} e^{ - i \phi } | + 1 \rangle + C_{0 \mu} | 0 \rangle
+ C_{-1, \mu} e^{ i \phi} | - 1 \rangle.
\label{eqn:psimu}
\end{align}
The quadrupole couplings between the three states are expressed as
\begin{align}
\alpha_{\mu \nu}^{(k)} = \sum_{m n} C_{m \mu} C_{n \nu} e^{ i (m - n) \phi }
\langle m | O_k | n \rangle~~(m, n = 1, 0, -1).
\end{align}
This leads to the matrix elements for quadrupole--strain coupling being written as
\begin{align}
& A_{\mu \nu} = \sum_k \alpha_{\mu \nu}^{(k)} A_{k, \varepsilon}
\nonumber \\
&~~~~~~
= \alpha_{\mu \nu}^{(u)} A_{u, \varepsilon}
+ \left[ \alpha_{\mu \nu}^{(v+)} \cos 2 \phi + i \alpha_{\mu \nu}^{(v-)} \sin 2 \phi \right]
A_{v,\varepsilon}
\nonumber \\
&~~~~~~~~~~~~
+ \left[ \alpha_{\mu \nu}^{(zx+)} \cos \phi + i \alpha_{\mu \nu}^{(zx-)} \sin \phi \right]
A_{zx, \varepsilon},
\label{eqn:Amunu}
\end{align}
where
\begin{align}
& \alpha_{\mu \nu}^{(u)} = \frac{1}{\sqrt{3}}
( C_{1 \mu} C_{1 \nu} - 2 C_{0 \mu} C_{0 \nu} + C_{-1, \mu} C_{-1, \nu} )
\nonumber \\
& \alpha_{\mu \nu}^{(v \pm)} = C_{1 \mu} C_{-1,\nu} \pm C_{-1, \mu} C_{1 \nu}
\nonumber \\
&  \alpha_{\mu \nu}^{(zx \pm)} = \frac{1}{\sqrt{2}}
\left[ ( C_{1 \mu} C_{0 \nu} \pm C_{0 \mu} C_{1 \nu})
- ( C_{0 \mu} C_{-1, \nu} \pm C_{-1, \mu} C_{0 \nu} ) \right].
\label{eqn:alpha-k}
\end{align}
\par

\subsection{Two-level system}
First, we consider an effective spin--strain interaction induced by an acoustic wave propagating in
the lattice in the subspace of $| \psi_1 \rangle$ and $| \psi_2 \rangle$.
As discussed above, this two-level system is coupled to time-dependent oscillating
strain fields such as $\varepsilon_{\lambda} = a_\lambda \cos \omega t$
($\lambda = 1, U_1, U_2$) with amplitude $a_\lambda$ and frequency $\omega$ of the acoustic
wave.
Here, we do not consider relative phase shifts (initial phase shifts at $t = 0$) between the three
strain components.
After calculating $\langle \psi_\mu | H_\varepsilon | \psi_\nu \rangle$ ($\mu, \nu = 1, 2$), we obtain
the two-level system coupled to the periodically time-dependent fields with the following form of
the effective Hamiltonian:
\cite{Koga20a,Matsumoto20,Koga20b,Shirley65,Son09}
\begin{align}
H_{\rm eff} (t) = \frac{1}{2}
\left(
\begin{array}{cc}
- \varepsilon_{12} - A_L \cos \omega t & A_T \cos \omega t \\
A_T^* \cos \omega t & \varepsilon_{12} + A_L \cos \omega t
\end{array}
\right).
\label{eqn:Ht}
\end{align}
Here, $\varepsilon_{12} = (\bar{E}_2 - \bar{E}_1) D_0$ is the energy difference between
$| \psi_1 \rangle$ and $| \psi_2 \rangle$.
For the quadrupole--strain interaction, the longitudinal ($A_L = A_{22} - A_{11}$) and transverse
($A_T = 2 A_{12}$) couplings are given by
$A_{\mu \nu} = \sum_k \alpha_{\mu \nu}^{(k)} A_k$ ($k = u,v,zx$) using Eqs.~(\ref{eqn:Amunu})
and (\ref{eqn:alpha-k}), where
\begin{align}
& A_u = g_a a_1,~~
A_v = ( g_b a_{U_1} + g_c a_{U_2} ) / \sqrt{3},
\nonumber \\
& A_{zx} = ( 2 g_d a_{U_1} - g_e a_{U_2} ) / \sqrt{6}.
\label{eqn:Ak}
\end{align}
Both $A_L$ and $A_T$ depend on $\theta$ and $\phi$.
The energy $\varepsilon_{12}$ remains a constant value during the rotation of the latter angle
$\phi$ of the magnetic field around the $z \parallel [111]$ axis.
\par

In a previous study, we investigated the $\phi$ dependence of the transition probability in the
two-level system for $B_z = 0$ ($\theta = \pi / 2$), changing $\varepsilon_{12}$ as a function of
only $\bar{b}$ (see Appendix~B).
\cite{Koga20b}
This is a great advantage for calculating the transition probability as a function of the field-rotation
angle $\phi$ for various values of $\varepsilon_{12}$.
In addition, the couplings $A_L$ and $A_T$ have compact forms owing to
$\alpha_{\mu \nu}^{(v+)} = \alpha_{\mu \nu}^{(zx-)} = 0$.
On the other hand, the acoustic-wave frequency $\omega$ is required to be energetically large
compared with the zero-field splitting $D$ to observe phonon-mediated resonance (for
simplicity, $\hbar = 1$ is used throughout the paper).
Here, $\varepsilon_{12} > D$ is given as a condition of energy-level splitting from
Eq.~(\ref{eqn:energy3}).
For $D / \omega > 1$, the single-phonon transition at $\varepsilon_{12} \simeq \omega$ cannot
be obtained when $\theta$ is fixed at $\pi /2$.
In addition, the second excited energy $\varepsilon_{13} = ( \bar{E}_3 - \bar{E}_1 ) D_0$ is not
large enough compared with $\varepsilon_{12}$, and the contribution from the quadrupole--strain
coupling with $| \psi_3 \rangle$ is not negligible. 
Therefore, $B_z$ should be finite ($\theta < \pi / 2$) for a lower frequency $\omega$.
Another advantage of a finite $B_z$ is that the strength of the spin--strain coupling $A_{zx}$
for the SQ transition can be evaluated from the $\phi$ dependence of the two-phonon transition
probability.
For $\theta = \pi / 2$, $A_{zx}$ is absent in Eqs.~(\ref{eqn:a11}) and (\ref{eqn:a22}), and it does
not appear in the longitudinal coupling $A_L = A_{22} - A_{11}$.
\par

In the present study, we change the field angle $\theta$ from $\pi / 2$ to zero, which is the simplest
way of changing the energy from a lower value ($\varepsilon_{12} / \omega < 1$) to a higher
value ($\varepsilon_{12} / \omega > 2$), where the magnetic field strength $B$ is fixed.
It is also necessary to adjust $B$ as well as $\theta$ to obtain $\varepsilon_{12} / \omega = 1$ and
$2$ for single- and two-phonon transition resonances, respectively.
As shown in Fig.~\ref{fig:1}(b), $\varepsilon_{12}$ is almost linearly dependent on $\theta$.
On the other hand, the second excited energy $\varepsilon_{13}$ is almost constant, maintaining
$( \varepsilon_{13} - \varepsilon_{12} ) / \omega > 2$ when $D_0 / \omega = 0.6$ and
$\gamma_e B / \omega = 2.0$.
To check the robustness of the two-level system describing the MAR in the spin-triplet system, we
also consider the influence from the second excited state
$| \psi_3 \rangle$ as well and extend Eq.~(\ref{eqn:Ht}) to the following $3 \times 3$ matrix form of
the three-level Hamiltonian:
\begin{align}
H_{\rm eff}^{\rm III} = E^{\rm III} + A^{\rm III} \cos \omega t.
\label{eqn:H3}
\end{align}
Here, the first term is written as 
\begin{align}
E^{\rm III} =
\left(
\begin{array}{ccc}
- ( \varepsilon_{12} / 2 ) - ( \varepsilon_{13} / 3 ) & 0 & 0 \\
0 & ( \varepsilon_{12} / 2 ) - ( \varepsilon_{13} / 3 ) & 0 \\
0 & 0 & 2 \varepsilon_{13} / 3
\end{array}
\right),
\end{align}
keeping the center of gravity of the three energy levels.
In the second term, the matrix elements of $A_{\mu \nu}^{\rm III}$  ($\mu, \nu = 1, 2, 3$) are given
as
\begin{align}
& A_{\mu \nu}^{\rm III} = \alpha_{\mu \nu}^{(u)} A_u
+ \left[ \alpha_{\mu \nu}^{(v+)} \cos 2 \phi + i \alpha_{\mu \nu}^{(v-)} \sin 2 \phi \right] A_v
\nonumber \\
&~~~~~~~~~~~~
+ \left[ \alpha_{\mu \nu}^{(zx+)} \cos \phi + i \alpha_{\mu \nu}^{(zx-)} \sin \phi \right] A_{zx},
\end{align}
using $A_k$ ($k = u, v, zx$) in Eq.~(\ref{eqn:Ak}).
\par

\section{Formulation}
\subsection{Floquet Hamiltonian}
A similar form of the time-dependent Hamiltonian in Eq.~(\ref{eqn:Ht}) has been frequently
studied by applying the Floquet theory.
According to Shirley's formulation,
\cite{Shirley65}
a problem of solving the time-dependent Schr\"{o}dinger equation is transformed to a
time-independent eigenvalue problem, which is described by an infinite-dimensional matrix form
of the Floquet Hamiltonian:
\begin{align}
\langle \alpha n | H_F | \beta m \rangle
= H_{\alpha \beta}^{ [ n - m ] } + n \omega \delta_{nm} \delta_{\alpha \beta}.
\end{align}
This comprises an infinite number of the Floquet states
$| \alpha n \rangle = | \alpha \rangle \otimes | n \rangle$, where $\alpha$ ($= \psi_1, \psi_2$)
denotes one of the two levels, and $n$ ($= 0, \pm 1, \pm2, \cdots$) is related to the
time-dependent wave function $e^{i n \omega t}$.
The block matrix $H^{[ n - m]}$ is only finite for $n - m = 0$ and $\pm1$ in the Floquet Hamiltonian
derived from Eq.~(\ref{eqn:Ht}).
The Floquet matrix is represented by
\begin{align}
H_F =
\left(
\begin{array}{ccccccc}
\ddots & ~ & ~ & \vdots & ~ & ~ & ~ \\
~ & H_{-2}^{[0]} & H^{[-1]} & {\bf 0} & {\bf 0} & {\bf 0} & ~ \\
~ & H^{[1]} & H_{-1}^{[0]} & H^{[-1]} & {\bf 0} & {\bf 0} & ~ \\
\cdots & {\bf 0} & H^{[1]} & H_0^{[0]} & H^{[-1]} & {\bf 0} & \cdots \\
~ & {\bf 0} & {\bf 0} & H^{[1]} & H_1^{[0]} & H^{[-1]} & ~ \\
~ & {\bf 0} & {\bf 0} & {\bf 0} & H^{[1]} & H_2^{[0]} & ~ \\
~ & ~ & ~ & \vdots & ~ & ~ & \ddots
\end{array}
\right),
\label{eqn:HFblock}
\end{align}
where the diagonal and off-diagonal sectors are given by
\begin{align}
H_n^{[0]} =
\left(
\begin{array}{cc}
- ( \varepsilon_{12} / 2 ) + n \omega & 0 \\
0 & ( \varepsilon_{12} / 2 ) + n \omega
\end{array}
\right),
\end{align}
\begin{align}
H^{[ \pm 1 ]} = \frac{1}{4}
\left(
\begin{array}{cc}
- A_L & A_T \\
A_T^* & A_L
\end{array}
\right),
\label{eqn:Hpm1}
\end{align}
respectively, and $H^{[n]} = \bf 0$ ($|n| \ge 2$) is the $2 \times 2$ zero matrix.
For the eigenvalue problem $H_F | q_\gamma \rangle = q_\gamma | q_\gamma \rangle$,
the $\gamma$th quasienergy $q_\gamma$ and the corresponding eigenfunction
$| q_\gamma \rangle$ are used for the following formula to calculate the time-averaged transition
probability between the $\alpha$ and $\beta$ states:
\cite{Shirley65,Son09}
\begin{align}
\bar{P}_{\alpha \rightarrow \beta} = \sum_m \sum_\gamma
| \langle \beta m | q_\gamma \rangle \langle q_\gamma | \alpha 0 \rangle |^2.
\label{eqn:P-form}
\end{align}
A similar formulation can also be applied to the three-level system including $| \psi_3 \rangle$,
and it is straightforward to rewrite the block parts $H_n^{(0)}$ and $H^{\pm 1}$ in $H_F$
as $3 \times 3$ matrix forms.
We choose $n$ in the range from $-50$ to $50$ in Eq.~(\ref{eqn:HFblock}) and check the
sufficient convergence of numerical results in the calculation of the multiphonon transition
probability.
\par

\subsection{Phonon absorption transition probability}
For the MAR, it is important to focus on the dependence of transition probability on the
field-rotation angle $\phi$ expected for the nearly degenerate Floquet states.
According to the Van Vleck perturbation theory, the time-averaged $n$-phonon transition
probability is obtained as
\cite{Son09}
\begin{align}
\bar{P}_{\psi_1 \rightarrow \psi_2}^{(n)}
= \frac{1}{2} \frac{ | v_{-n} |^2 }{ ( n \omega - \varepsilon_{12} + 2 \delta_n )^2 / 4 + | v_{-n} |^2 },
\label{eqn:Pn}
\end{align}
where $ - \varepsilon_{12} / 2 \simeq \varepsilon_{12} / 2 - n \omega$ between the two
Floquet states, for instance, $| \psi_1,0 \rangle$ and $| \psi_2, - n \rangle$.
Equation (\ref{eqn:Pn}) is valid for $n = 1$ and $2$ when the effective transverse coupling
$v_{-n}$ between the two states is calculated up to the first-order perturbation of $A_T$ as
\cite{Matsumoto20,Koga20b}
\begin{align}
v_{-n} = - \frac{1}{2} n \omega J_{-n} \left( \frac{ A_L }{ \omega } \right) \frac{ A_T }{ A_L }.
\label{eqn:v-n}
\end{align}
Here, the $k$th Bessel function of the first kind, $J_k$, is used for $A_L$.
For the energy shift $\delta_n$, the leading term is represented by
\begin{align}
\delta_n = - \sum_{k \ne - n} \frac{ | v_k |^2 }{ \varepsilon_{12} + k \omega }.
\label{eqn:deltan}
\end{align}
In the weak coupling limit ( $| A_L |, | A_T | \ll \omega$ ),
\begin{align}
| v_{-n} | \simeq \frac{ | A_T | }{ 2^{n+1} ( n - 1 )! } \left( \frac{ | A_L | }{ \omega } \right)^{n-1}~~
(n \ge 1),
\label{eqn:v-nw}
\end{align}
which leads to simple analytic forms of the transition probability at $\varepsilon_{12} =  \omega$
and $2 \omega$ as
\cite{Matsumoto20,Koga20b}
\begin{align}
& \bar{P}_{\psi_1 \rightarrow \psi_2}^{(1)} (\varepsilon_{12} = \omega)
= \frac{1}{2} \frac{1}{ 1 + [ | A_T | / (8 \omega) ]^2 },
\label{eqn:P1} \\
& \bar{P}_{\psi_1 \rightarrow \psi_2}^{(2)} (\varepsilon_{12} = 2 \omega)
= \frac{1}{2} \frac{ 1 }{ 1 + (4/9) ( | A_T | / | A_L | )^2 },
\label{eqn:P2}
\end{align}
for finite $| A_T |$, respectively.
The $\phi$ dependence of the transition probability is derived from the change in $A_L$ and
$A_T$ under the field rotation around the $z \parallel [111]$ axis.
Equations~(\ref{eqn:P1}) and (\ref{eqn:P2}) show that $\bar{P}^{(1)}$ ($\simeq 1/2$) for small
$| A_T |$, while $\bar{P}^{(2)}$ decreases from the maximum $1 / 2$ for $| A_T | \simeq | A_L |$
and vanishes as $A_L \rightarrow 0$.
We note that the longitudinal coupling $A_L$ is required for the latter two-phonon transition
process in addition to the transverse coupling $A_T$.
\par

Experimentally, multiphonon transition processes have been detected by photoluminescence
excitation spectroscopy for the NV center.
\cite{Chen18}
The observed phonon sidebands derived from the NV orbital transitions were explained by
quantum transitions between multiphonon dressed states.
In this case, a photon field is responsible for the transverse coupling between two sublevels
of electronic orbital states.
As described in Sect.~3.1, the $n$-phonon dressed state is represented by the Floquet state
$| \alpha n \rangle$ or $| \beta n \rangle$ ($n = 0, \pm 1, \pm 2, \cdots$), where $\alpha$ and
$\beta$ denote the two levels.
For instance, $| \alpha 0 \rangle$ and $| \beta, -n \rangle$ are coupled in the $n$-phonon
absorption transition associated with longitudinal phonon coupling between the two levels.
\par

\subsection{Quantum transition with two-phonon absorption}
\begin{figure}
\begin{center}
\includegraphics[width=7cm,clip]{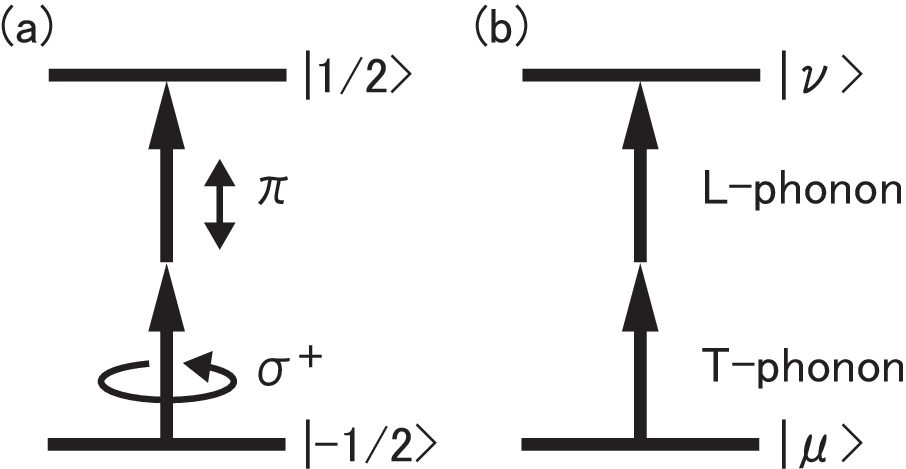}
\end{center}
\caption{
(a) Two-photon transition.
Right-hand circularly polarized and linearly polarized photons ($\sigma^+$ and $\pi$) are
absorbed.
\cite{Gromov00}
(b) Two-phonon transition.
One absorbed phonon is for transverse coupling $A_T$ ($T$-phonon) and the other is for
longitudinal coupling $A_L$ ($L$-phonon) in the two levels.
}
\label{fig:2}
\end{figure}
\begin{figure}
\begin{center}
\includegraphics[width=7cm,clip]{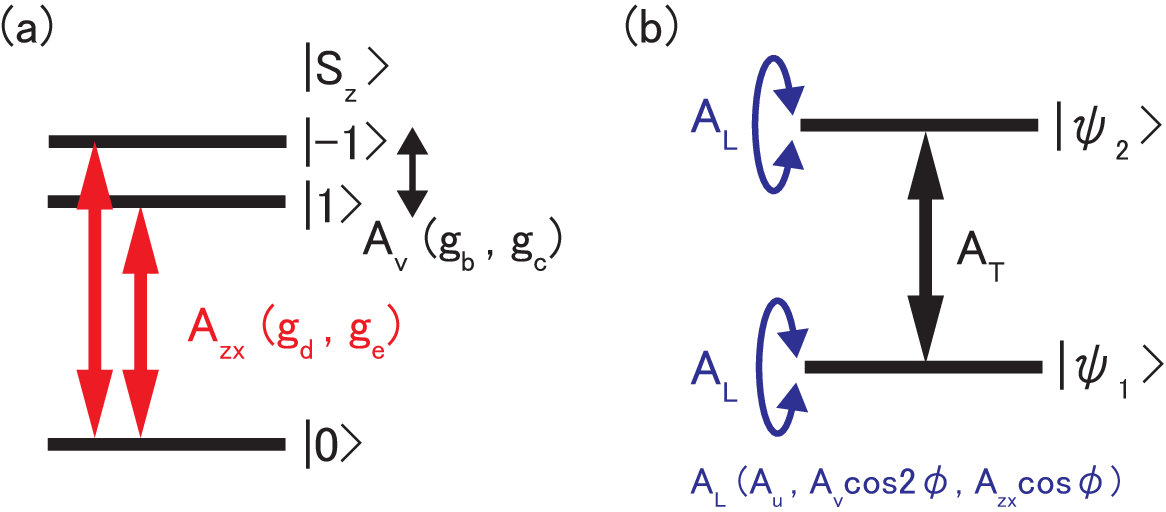}
\end{center}
\caption{
(Color online)
(a) Spin--strain couplings $A_v$ for the DQ transition and $A_{zx}$ for the SQ transition
(diagonal coupling $A_u$ is not shown).
As given in Eq.~(\ref{eqn:Ak}), $A_v$ contains the coupling-strength parameters ($g_b, g_c$), and
$A_{zx}$ contains ($g_d, g_e$).
The latter is focused on in the present study.
(b) Transverse ($A_T$) and longitudinal ($A_L$) couplings in the two-level system.
The latter is represented by the linear combination of $A_u$, $A_v \cos 2 \phi$, and
$A_{zx} \cos \phi$ with respect to the field-rotation angle $\phi$ in Eq.~(\ref{eqn:A-L}).
}
\label{fig:3}
\end{figure}
In two-level systems with $S = 1/2$ spins, a two-photon transition is induced by circularly
polarized and linearly polarized photon fields simultaneously between the
$| S_z = \pm 1/2 \rangle$ states in a static magnetic field.
\cite{Gromov00}
As shown in Fig.~\ref{fig:2}(a), it requires not only a right-hand circularly polarized $\sigma^+$
photon but also a linear polarized $\pi$ photon in the transition
$| -1/2 \rangle \rightarrow | 1/2 \rangle$
(left-hand circularly polarized $\sigma^-$ plus $\pi$ in $| 1/2 \rangle \rightarrow | -1/2 \rangle$).
For two sublevels of the $S = 1$ system in Fig.~\ref{fig:2}(b), there is the following correspondence
between the two-photon and two-phonon transition processes:
$\sigma^\pm$-photon $\leftrightarrow$ $T$-phonon and $\pi$-photon $\leftrightarrow$
$L$-phonon.
\par

In the two-level model discussed here, three different spin--strain coupling components $A_k$
($k = u, v, zx$) are involved in $A_L$ ($= A_{22} - A_{11}$) and $A_T$ ($= 2 A_{12}$) as
\begin{align}
& A_L = \alpha_-^{(u)} A_u + \alpha_-^{(v+)} A_v \cos 2 \phi + \alpha_-^{(zx+)} A_{zx} \cos \phi
\label{eqn:A-L} \\
& A_T = 2 [ \alpha_{12}^{(u)} A_u + \alpha_{12}^{(v+)} A_v \cos 2 \phi
+ \alpha_{12}^{(zx+)} A_{zx} \cos \phi]
\nonumber \\
&~~~~~~~~
+ i 2 [ \alpha_{12}^{(v-)} A_v \sin 2 \phi + \alpha_{12}^{(zx-)} A_{zx} \sin \phi ],
\end{align}
where $\alpha_-^{(k)} = \alpha_{22}^{(k)} - \alpha_{11}^{(k)}$ ($k = u, v+, zx+$) in
Eqs.~(\ref{eqn:Amunu}) and (\ref{eqn:alpha-k}).
The couplings $A_v$ and $A_{zx}$ are associated with the DQ and SQ transitions, respectively,
in Fig.~\ref{fig:3}(a).
The latter contains the spin--strain coupling-strength parameters $g_d$ and $g_e$ to be
evaluated from the longitudinal coupling $A_L$ in Fig.~\ref{fig:3}(b) in the present analysis.
We note that each coupling $A_k$ depends on the three strain-field components
$\varepsilon_\lambda$  ($\lambda = 1, U_1, U_2$) in Eq.~(\ref{eqn:Ak-ep}).
We can obtain a field-rotation angle $\phi$ for $A_L \rightarrow 0$ when we choose a single
component $\varepsilon_{U_1}$ (elongation strain) or the other two components $\varepsilon_1$
and $\varepsilon_{U_2}$ (shear strain) together.
For surface acoustic waves with small decay, we can consider that the latter two components are
dominant, although $\varepsilon_{U_1}$ with a different phase shift is generated by the
penetration effect (see Appendix~C).

\section{Results}
\subsection{Two-phonon transition probability for weak coupling}
It is important to find the values of $\phi$ for a significant decrease in the two-phonon transition
probability in Eq.~(\ref{eqn:P2}), which is useful for evaluating the spin--strain coupling strength.
When the field strength $B$ is fixed, $\phi = \phi_B$ at $A_L \rightarrow 0$ is determined from
Eq.~(\ref{eqn:A-L}).
Using the ratios of spin--strain coupling strength $a_u = A_u / ( \sqrt{3} A_v )$ and
$a_{zx} = A_{zx} / A_v$, the condition of $A_L \rightarrow 0$ is reduced to
\begin{align}
\gamma_B^{(u)} a_u = \cos 2 \phi_B + \gamma_B^{(zx)} a_{zx} \cos \phi_B,
\label{eqn:a-k}
\end{align}
where $\gamma_B^{(u)}$ and $\gamma_B^{(zx)}$ are defined as
\begin{align}
\gamma_B^{(u)} = - \frac{ \sqrt{3} \alpha_{B -}^{(u)} }{ \alpha_{B -}^{(v+)} },~~
\gamma_B^{(zx)} = \frac{ \alpha_{B -}^{(zx+)} }{ \alpha_{B -}^{(v+)} }~~
( \alpha_{B -}^{(k)} \equiv \alpha_{B, 22}^{(k)} - \alpha_{B, 11}^{(k)} ).
\end{align}
The values of $\alpha_{B, \mu \mu}^{(k)}$ ($k = u, v+, zx+$) in Eq.~(\ref{eqn:Amunu}) are
calculated for $\varepsilon_{12} = 2 \omega$, and this energy splitting is adjusted by changing
$B$ and $\theta$.
Let us consider that we obtain $\phi_{B 1}$ and $\phi_{B 2}$ in
$0 < \phi_{B 1}, \phi_{B 2} < \pi / 2$ from the data of two-phonon transition probabilities for
$(B_1, \theta_1)$ and $(B_2, \theta_2)$, respectively.
Using Eq.~(\ref{eqn:a-k}), the solutions of $a_{zx}$ and $a_u$ are given by the simultaneous
equations for $B_1$ and $B_2$ as
\begin{align}
& a_{zx} = \frac{ \gamma_{B 1}^{(u)} \cos 2 \phi_{B 2}
- \gamma_{B 2}^{(u)} \cos 2 \phi_{B 1} }
{ \gamma_{B 2}^{(u)} \gamma_{B 1}^{(zx)} \cos \phi_{B 1}
- \gamma_{B 1}^{(u)} \gamma_{B 2}^{(zx)} \cos \phi_{B 2} },
\label{eqn:a-zx} \\
& a_u = \frac{ \gamma_{B 1}^{(zx)} \cos \phi_{B 1} \cos 2 \phi_{B 2}
- \gamma_{B 2}^{(zx)} \cos \phi_{B 2} \cos 2 \phi_{B 1} }
{ \gamma_{B 2}^{(u)} \gamma_{B 1}^{(zx)} \cos \phi_{B 1}
- \gamma_{B 1}^{(u)} \gamma_{B 2}^{(zx)} \cos \phi_{B 2} }.
\label{eqn:a-u}
\end{align}
\par

\begin{figure}
\begin{center}
\includegraphics[width=7cm,clip]{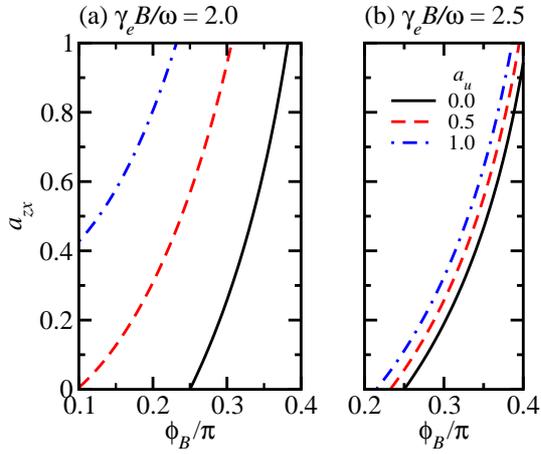}
\end{center}
\caption{
(Color online)
Coupling-strength ratio $a_{zx}$ as a function of $\phi_B$ ($A_L \rightarrow 0$) in
Eq.~(\ref{eqn:a-k}) plotted for various values of $a_u$.
Here, $a_u = 0.0, 0.5$, and $1.0$.
(a) $\gamma_e B / \omega = 2.0$, where the coefficients $\gamma_B^{(u)} = 1.638$ and
$\gamma_B^{(zx)} = 2.040$ are the values for $\varepsilon_{12} = 2 \omega$.
(b) $\gamma_e B / \omega = 2.5$, where $\gamma_B^{(u)} = 0.219$ and
$\gamma_B^{(zx)} = 2.777$ for $\varepsilon_{12} = 2 \omega$.
}
\label{fig:4}
\end{figure}

In Fig.~\ref{fig:4}, we show how $a_{zx}$ and $a_u$ depend on the measured values of $\phi_B$
in Eq.~(\ref{eqn:a-k}).
Here, $a_{zx}$ is plotted as a function of $\phi_B$ for various values of $a_u$ for different
magnetic field strengths:
(a) $\gamma_e B / \omega = 2.0$ and (b) $\gamma_e B / \omega = 2.5$.
Assuming that $| a_{zx} |, | a_u | \ll 1$, $\phi_B / \pi = 1/4$ is obtained for arbitrary values of $B$.
The increasing shift of $\phi_B$ depends on the increase in $a_{zx}$ and the decrease in $a_u$.
In Fig.~\ref{fig:4}(b), the shift of $\phi_B$ is not greatly dependent on the change in $a_u$
because of the very small value of $\gamma_B^{(u)} \simeq 0.2$.
This indicates that we can estimate the magnitude of $a_{zx}$ to some extent from the measured
value of $\phi_B$ regardless of $a_u$.
We also predict that for an extremely small $| a_{zx} |$, $\phi_B$ can be found in the range of
$0.2 < \phi_B / \pi < 0.3$ in Fig.~\ref{fig:4}(b), where $| a_u | \lesssim 1$ is assumed.
In the next subsection, we demonstrate how to determine $\phi_B$ from the calculated
transition probabilities.
\par

\subsection{Resonance in transition probability}
\begin{figure}
\begin{center}
\includegraphics[width=6cm,clip]{fig5top.eps} \\
\vspace*{-0.7cm}
\includegraphics[width=7cm,clip]{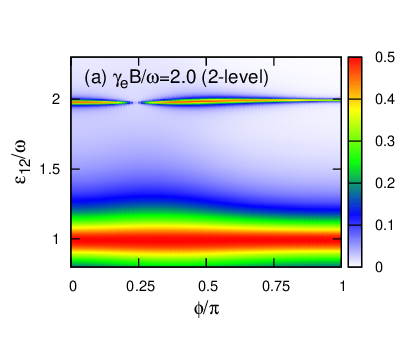} \\
\vspace*{-1.0cm}
\includegraphics[width=7cm,clip]{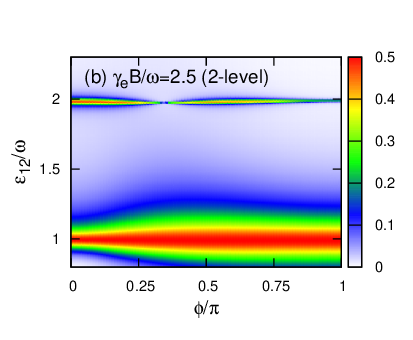}
\end{center}
\caption{
(Color online)
Contour maps of $\bar{P}_{ \psi_1 \rightarrow \psi_2 }$ as a function of $\phi$ and
$\varepsilon_{12} / \omega$ ($ > 0.8$) for the two-level system, where $A_{13} = A_{23} = 0$ for
the quadrupole--strain couplings with $| \psi_3 \rangle$.
At the top, $\varepsilon_{12} / \omega$ is shown as a function of the field direction $\theta$.
The data $\bar{P}$ are plotted for $D_0 / \omega = 0.6$ and
$(A_u  / \sqrt{3}, A_v, A_{zx}) / \omega= (0.1, 0.2, 0.1)$ under magnetic fields of
(a) $\gamma_e B / \omega = 2.0$ and (b) $\gamma_e B / \omega = 2.5$.
}
\label{fig:5}
\end{figure}
We discuss the two-level system where the contribution from the second excited
state $| \psi_3 \rangle$ is ignored.
Using Eq.~(\ref{eqn:P-form}), we calculate the time-averaged transition probability $\bar{P}$
between the lowest-lying states $| \psi_1 \rangle$ and $| \psi_2 \rangle$.
Here, the two-level splitting $\varepsilon_{12}$ is changed as a function of $\theta$, keeping the
field strength $B$ indicated in Fig.~\ref{fig:1}(b).
For our purpose, $\varepsilon_{12}$ must be in $1 \le \varepsilon_{12} / \omega \le 2$ at least.
The contour maps of $\bar{P} (\phi, \varepsilon_{12})$ are shown in Fig.~\ref{fig:5} to capture
typical features of the field-angle-resolved MAR.
\par

One of the pronounced features is a broad peak band lying at
$\varepsilon_{12} / \omega = 1$ for all values of $0 \le \phi / \pi \le 1$.
This is because of the single-phonon transition process.
The peak width of $\bar{P} (\varepsilon_{12} / \omega)$ is mainly dependent on
$| A_T |$ ($= 2 | A_{12} |$) related to $\delta_1$ for weak coupling in Eqs.~(\ref{eqn:deltan})
and (\ref{eqn:v-nw}).
On the other hand, a narrow peak line at $\varepsilon_{12} / \omega \simeq 2$ is also marked,
which corresponds to the two-phonon transition process.
There exists a node in this $\phi$-dependent $\bar{P}$ owing to $A_L \rightarrow 0$.
This is important for evaluating $A_k$ ($k = u, v, zx$) for the spin--strain coupling between the two
levels.
In fact, the value of $\phi$ for $\bar{P} (\varepsilon_{12} / \omega \simeq 2) \rightarrow 0$
is shifted by adjusting the field direction $\theta$ for $\varepsilon_{12} / \omega \simeq 2$
after changing the field strength $B$.
In Fig.~\ref{fig:5}(a), $\bar{P}$ exhibits a node at $\phi / \pi = 0.24$ for
$( \gamma_e B / \omega, \theta / \pi ) = (2.0,0.247)$, which moves toward 0.5 with an increase in
$B$.
This is confirmed in Fig.~\ref{fig:5}(b), where the node of $\bar{P}$ shifts to $\phi / \pi = 0.35$
for $( \gamma_e B / \omega, \theta / \pi ) = (2.5,0.199)$.
The $\bar{P}$-node angles in Figs.~\ref{fig:5}(a) and \ref{fig:5}(b) correspond to the values of
$\phi_B$ for $a_u = A_u / ( \sqrt{3} A_v ) = 0.5$ and $a_{zx} = A_{zx} / A_v = 0.5$ in
Figs.~\ref{fig:4}(a) and \ref{fig:4}(b), respectively.
Both figures also show that $\phi_B$ shifts toward $\pi / 2$ with an increase in $a_{zx}$ and a
decrease in $a_u$.
For a relatively large $\gamma_e B / \omega \simeq 2.5$, $\phi_B$ is not greatly dependent on
$a_u$, as shown in Fig.~\ref{fig:4} (b), which allows a rough estimate of $a_{zx}$.
Thus, the field-rotation angles $\phi_{B 1}$ and $\phi_{B 2}$ required for Eqs.~(\ref{eqn:a-zx}) and
(\ref{eqn:a-u}) can be obtained from experimentally observed transition probabilities for different
magnetic fields $B_1$ and $B_2$, respectively.
\par

In the above argument so far, the contribution from $| \psi_3 \rangle$ has not been considered in
calculating the transition probability.
Under a finite magnetic field, the level splitting $\varepsilon_{13}$ is not very large compared with
$\varepsilon_{12}$ in the $S = 1$ triplet, for instance,
$\varepsilon_{13} \simeq 2 \varepsilon_{12}$ for
$( \gamma_e B / \omega, \theta / \pi ) \simeq (2.0,0.3)$ in Fig.~\ref{fig:1}(b).
Although the transitions between two lowest-lying levels can be affected by the quadrupole--strain
couplings $A_{13}$ and $A_{23}$ with $| \psi_3 \rangle$ in the $S = 1$ triplet, we show that the
$| \psi_3 \rangle$ contribution is a minor effect on the two-phonon transition probability at
$\phi \simeq \phi_B$ for $A_L \rightarrow 0$ in Appendix~D.
\par

\subsection{Application to NV spin}
Here, the above argument is applied to the electron $S = 1$ triplet state of the NV center having
the zero-field splitting $D = 2.87$ GHz.
This indicates that $D_0 / \omega = D / (3\omega) = 0.6$ is used assuming that
$\omega / 2 \pi = 1.6$ GHz for the frequency of an oscillating strain field.
From a recent experiment, the ratio of the spin--strain coupling parameters has been found to be
$g_d / g_b = 0.5 \pm 0.2$ ($g_b g_d > 0$ is assumed here) associated with uniaxial
stress fields.
\cite{Chen20}
As described in Appendix~A, it corresponds to $- d / (\sqrt{2} b)$ in Eq.~(\ref{eqn:SQ-DQ}) for the
spin--stress coupling strength:
$d$ is related to $\{ g_{25}, g_{26} \}$ for the SQ transition and $b$ is related to
$\{ g_{15}, g_{16} \}$ for the DQ transition in Eq.~(\ref{eqn:spin-stress}).
In fact, this value of $g_d / g_b$ is about ten times larger than a numerical result based on density
functional theory,
\cite{Udvarhelyi18}
and the coupling ratio is highly desirable to be quantified.
\par

Using Eq.~(\ref{eqn:Ak}), the ratio $a_{zx} = A_{zx} / A_v$ is related to the spin--strain
coupling-strength parameters as
\begin{align}
a_{zx} = \frac{ 2 g_d a_{U_1} - g_e a_{U_2} }{ \sqrt{2} ( g_b a_{U_1} + g_c a_{U_2} ) },
\end{align}
which is reduced to $a_{zx} = \sqrt{2} g_d / g_b$ for the strain amplitude $a_{U_1}$ of a uniaxial
strain field ($a_{U_2} = 0$ in the absence of a shear strain field).
This strain can be induced by a longitudinal acoustic wave propagating in the direction of the
$Z$ axis ($\varepsilon_{U_1} = 2 \varepsilon_{ZZ} / \sqrt{3}$).
Let us substitute the above experimentally measured value $\sqrt{2} g_d / g_b$ for $a_{zx}$ in
Eq.~(\ref{eqn:a-k}) when the magnetic field $\gamma_e B / \omega = 2.5$ is applied.
Assuming that $| \gamma_B^{(u)} a_u | = | \gamma_B^{(u)} A_u / ( \sqrt{3} A_v ) | \ll 1$, we can
estimate $0.37 < \phi_B / \pi < 0.40$ as a possible observed field-rotation angle at which the
two-phonon transition probability is significantly decreased.
For a smaller value of $a_{zx}$, $\phi_B / \pi$ shifts toward $0.25$ (for instance, $a_{zx} = 0.1$
gives $\phi_B / \pi = 0.28$).
\par

For another possible measurement, $a_{zx}$ is reduced to $- g_e / (\sqrt{2} g_c)$ using
a shear strain field, for instance, $\varepsilon_{YZ} = \varepsilon_{ZX}$ and
$\varepsilon_{XY} = 0$ with finite amplitude $a_{U_2}$ (and $a_{U_1} = 0$).
This strain field can be induced by a bulk transverse acoustic wave propagating in the $[110]$
direction of the cubic ($XYZ$) frame.
For the transverse displacement $u_Z$, a simple plane-wave representation is given by
$q \rightarrow 0$ in Eq.~(\ref{eqn:SAW}).
In Eq.~(\ref{eqn:SQ-DQ}), the ratio $g_e / g_c$ equals $\sqrt{2} e / c$.
For the latter, $e$ is related to the spin--stress couplings $\{ g_{25},  g_{26} \}$ for the SQ transition
and $c$ is related to $\{ g_{15}, g_{16} \}$ for the DQ transition in Eq.~(\ref{eqn:spin-stress}).
\par

Finally, we comment on the difference between our method and the previous theoretical proposal
for optical measurements using static stress fields in Ref.~5.
According to the method in Ref.~5, $g_{25}$ and $g_{26}$ are measured as the
coupling-strength parameters of an effective transverse field that brings about the electronic
ground-state level anticrossing.
This is due to the mixing of the singlet ground state $| 0 \rangle$ and one of the spin doublet
states $| \pm 1 \rangle$, which can be realized by applying a strong magnetic field
$\simeq 1024$ G energetically comparable to the zero-field splitting.
As a result of the level anticrossing, these parameters can be evaluated from the Larmor
frequency $f_L$ in the photoinduced spin dynamics as
$| g_{25} |, | g_{26} | \propto f_L / | \sigma |$, where $| \sigma |$ is the strength of a static and
uniaxial stress field generating the corresponding spin--stress interaction.
This indicates that the application of a sufficiently strong stress is required for a measurable $f_L$
of a megahertz order if $| g_{25} |$ and $| g_{26} |$ are extremely small.
For the NV center,  the experimentally measured coupling-strength parameters $a_2$, $b$, and
$c$ in Eq.~(\ref{eqn:spin-stress}) are on the order of a few megahertz per GPa.
The coupling-strength parameters $d$ and $e$ related to $g_{25}$ and $g_{26}$ have been
considered to be much smaller in magnitude ($| d / b | \simeq 0.06$ and $| e / c | \simeq 0.2$ from
Ref.~5).
\par

In our method, on the other hand, we target the ratios $g_d / g_b$ (or $d / b$) and $g_e / g_c$
(or $e / c$) to compare the couplings between the SQ and DQ transitions driven by an acoustic
wave.
In principle, the coupling-strength ratios can be evaluated from the two-phonon transition
probabilities, which are dependent not on the amplitudes of oscillating strain (or stress) fields but
on the rotation angle of an applied magnetic field.
This is an advantage of our method for the measurement of small coupling-strength parameters
and is also applicable to a wide range of spin--strain coupling strength.
Except for technical problems in the observation of the two-phonon absorption, it is worthwhile to
elucidate the potential of MAR measurements presented here for future applications to the NV
centers in diamond.
\par

We also comment on an off-axis magnetic field effect on the NV spin.
In Fig.~\ref{fig:5}(a), we adopted $( \gamma_e B / \omega, \theta / \pi ) = (2.0, 0.247)$ to
obtain the two-phonon transition probabilities at $\varepsilon_{12} / \omega \simeq 2$ for
$D_0 / \omega = 0.6$, and $\omega / 2 \pi = 1.6$ GHz for the NV center, as mentioned above.
The values of $\omega$ and $\theta$ lead to $B \simeq 1100$ G and the transverse field
component $B_\bot = B \sin \theta \simeq 800$ G.
This value of $B_\bot$ exceeds $200$ G, at which a signal becomes invisible in optically detected
electron spin resonance spectra.
\cite{Tetienne12}
For the MAR measurements proposed here, an abrupt decrease in the two-phonon transition
probability between $| \psi_1 \rangle$ and $| \psi_2 \rangle$ becomes more prominent with
increasing $B$, where $\theta$ is adjusted for $\varepsilon_{12} / \omega \simeq 2$ and $B_\bot$
itself does not affect the results.
In the present study, we have confirmed this by investigating the contribution from
$| \psi_3 \rangle$ in the third level of the NV spin, as discussed in Appendix~D.

\subsection{Analogy between strain and electric-field effects}
Note also that the NV spin can couple to an electronic field $\bE = (E_x, E_y, E_z)$ as well as
a strain field.
According to the group-theoretical analysis,
\cite{Matsumoto17}
the components of a spin-dependent electric dipole $\bp = (p_x, p_y, p_z)$ can be described by
the second-rank tensors of spin operators, and the interaction Hamiltonian $H_E = - \bp \cdot \bE$
for the $C_{3v}$ symmetry is given by
\cite{Udvarhelyi18,Matsumoto17}
\begin{align}
H_E = K_0 E_z O_u + K_1 ( E_x O_{zx} + E_y O_{yz} ) + K_2 ( - E_x O_v + E_y O_{xy} ).
\label{eqn:HE}
\end{align}
Here, $O_u$, $O_v$, and $O_{zx}$ are the same as the quadrupole operators in
Eq.~(\ref{eqn:Ok}), and $\{ O_{yz}, O_{xy} \} = \{ S_y S_z + S_z S_y, S_x S_y + S_y S_x \}$.
The coefficients $K_l$ ($l = 0, 1, 2$) are related to the quadrupole coupling
$\langle m | O_k | n \rangle$ ($l = | m - n |$) for $k = u, v, yz, zx$, and $xy$, where $m$ and $n$
represent one of the $S_z$ components.
To apply the above argument about the spin--strain interaction for the NV spin, we choose a
finite electric field in the $x$ direction ($E_y = E_z =0$).
In comparison with the interaction Hamiltonian $H_\varepsilon$ in Eq.~(\ref{eqn:Hep}), we obtain
$H_E$ by replacing the strain-dependent coefficient $A_{k, \varepsilon}$ as
\begin{align}
A_{v, \varepsilon} \rightarrow - K_2 E_x,~~
A_{zx,\varepsilon} \rightarrow K_1 E_x,
\end{align}
and $A_{u, \varepsilon} = 0$.
\par

As discussed in Sect.~2.3, we consider that $E_x$ is periodically time-dependent and coupled to
the two-level system in the subspace of $| \psi_1 \rangle$ and $| \psi_2 \rangle$ when the static
magnetic field $B$ is applied.
For the two-photon transition probability $\bar{P}^{(2)}$, a significant decrease in $\bar{P}^{(2)}$
is also expected at field-rotation angle $\phi_B$ for the vanishment of the longitudinal coupling
$A_L$, as in the case of phonons in Eq.~(\ref{eqn:P2}).
The condition for $A_L \rightarrow 0$ is given by Eq.~(\ref{eqn:a-k}), where $a_u = 0$
and $a_{zx} = K_1 / ( - K_2 )$ for $E_x$.
This indicates that the existence of $K_1$ can be confirmed by finding a node at $\phi_B$ in a
narrow line of resonance peak $\bar{P} (\phi)$ at $\varepsilon_{12} / \omega \simeq 2$.
Indeed, $K_1$ has been considered to be negligibly small so far, and an upper limit
$| a_{zx} | < 0.047$ has been reported by a recent experimental study of the NV spin,
\cite{Michl19}
implying $\phi_B \simeq \pi / 4$.
However, an ac electric field $E_x$ may cause induction of a magnetic field perpendicular to the
direction of $\bE$.
This magnetic field is also coupled to the spin states and inevitably disturbs the pure electric-field
effect.
\par

Therefore, both magnetic and electric fields must be considered for such dynamic coupling with
a spin state.
For our purpose, a linearly polarized microwave characterized by $E_x$ and $B_z$ is a good
candidate for probing the field-rotation angle $\phi_B$ for $A_L \rightarrow 0$.
The corresponding interaction Hamiltonian is given by
\begin{align}
H_{\rm MW} = K_1 E_x O_{zx} - K_2 E_x O_v - \gamma_e B_z S_z.
\end{align}
The matrix elements of the third term,
\begin{align}
\alpha_{B, \mu \nu}^{(z)} = \langle \psi_\mu | S_z | \psi_\nu \rangle
= C_{1 \mu} C_{1 \nu} - C_{-1, \mu} C_{-1, \nu}~~(\mu, \nu = 1,2),
\end{align}
lead to a finite $\gamma_B^{(u)}$ in Eq.~(\ref{eqn:a-k}), where $a_u$ and $\gamma_B^{(u)}$ are
replaced as
\begin{align}
a_u \rightarrow \frac{ \gamma_e B_z }{ K_2 E_x },~~
\gamma_B^{(u)} \rightarrow \frac{ \alpha_{B -}^{(z)} }{ \alpha_{B -}^{(v+)} },
\end{align}
and $\alpha_{B -}^{(z)} = \alpha_{B, 22}^{(z)} - \alpha_{B, 11}^{(z)}$.
Under the assumption that $a_{zx} = K_1 / ( - K_2 ) = 0$, a finite $\phi_B$
($0 < \phi_B < \pi / 2$)  for $A_L \rightarrow 0$ can be obtained as
$\phi_B = (1/2) \arccos [ \gamma_B^{(u)} a_u ]$ for $| \gamma_B^{(u)} a_u | < 1$.
For a finite $a_{zx}$, $\phi_B$ can be determined from the $\phi$ dependence of two-photon
transition probabilities for various values of $B$.
Using Eqs.~(\ref{eqn:a-zx}) and (\ref{eqn:a-u}), we can evaluate $a_{zx}$ and $a_u$ from two
values of $\phi_B$ ($\phi_{B 1}$ and $\phi_{B 2}$).
Thus, a finite value of the evaluated $a_{zx}$ confirms that $K_1$ in Eq.~(\ref{eqn:HE}) should be
considered as a relevant coupling parameter.

\section{Conclusion}
In this paper, we demonstrated how to evaluate coupling-strength parameters in the
spin--strain interaction for the NV spin-1 states, showing the applicability of field-angle-resolved
MAR as a new probe of acoustically driven quantum spin transitions.
Floquet theory was applied to describe the phonon-driven transitions by longitudinal ($A_L$) and
transverse ($A_T$) quadrupole--strain couplings between the two lowest-lying levels, which were
investigated by rotating a magnetic field perpendicular to the NV axis in our previous study.
\cite{Koga20b}
In the present study, the level splitting was adjusted by changing the field direction $\theta$
relative to the NV axis to use a lower resonance frequency in ultrasonic measurements.
This was also important for the evaluation of the spin--strain coupling strength
from the two-phonon transition probability, which we had not taken into account previously.
\cite{Koga20b}
Here, we showed the important role of coupling $A_L$ in the dependence of the
two-phonon transition probability on the field-rotation angle $\phi$.
It was crucial to find a significant decrease in the $\phi$ dependence owing to $A_L \rightarrow 0$
and evaluate the spin--strain coupling strength associated with the SQ transition
(in Fig.~\ref{fig:3}, $g_d$ and $g_e$ in $A_{zx}$).
In addition to lowering the excitation energies, it was also important, for this evaluation, to tilt the
field direction from the $\theta = \pi / 2$ plane.
Considering all of the three $S = 1$ states, we confirmed the robustness of the two-level
treatment.
\par

In the last decade, much experimental effort has been made to realize the perfect alignment of NV
defects oriented along one of the threefold axes, [111], in diamond crystals.
\cite{Michl14,Lesik14,Miyazaki14,Ozawa17,Mizuno21,Tatsuishi21}
By adjusting the specific orientation, the signals from the dominantly aligned NV defects can be
intensified selectively among the defects oriented along the four possible threefold axes in
electron spin resonance measurements.
This is of great advantage to the MAR measurements presented here to enable control by rotating
a magnetic field around the [111] axis.
Applying a magnetic field perpendicular to a specific NV axis is also important for probing the
quantum transitions between the NV spin states, as revealed by a recent experimental study of
optically detected magnetic resonance.
\cite{Yamaguchi20}
These latest developments will be driving forces for testing the efficiency of ultrasonic
measurements to probe acoustically driven quantum spin transitions.
\par

The idea of field-angle-resolved MAR can be developed to investigate other defect spins coupled
to strain fields at the atomic scale.
For instance, silicon vacancies in diamond
\cite{Sohn18,Meesala18,Maity20}
and in silicon carbide (SiC)
\cite{Whiteley19a,Soltamov19,Whiteley19b}
are also leading platforms for spin--strain coupling with high sensitivity.
Recently, acoustically driven spin resonance measurements have been applied to the defect
spins in SiC described by the spin-$3/2$ system and have shown that phonon-driven SQ and DQ
resonance transitions can be changed by rotating a static in-plane magnetic field on the SiC
surface.
\cite{Hernandez-Minguez20}
The unusual field-direction dependence observed in single-phonon transition probabilities is
an intriguing issue, and we should generalize the present analysis for practical measurements
using a surface acoustic wave.
A consistent study of multiphonon-driven quantum spin transitions is also strongly urged as
counterparts of multiphoton absorption transitions between spin levels, as observed by recent
optical measurements of silicon vacancies.
\cite{Singh22}

\bigskip
{\footnotesize
{\bf Acknowledgments}~~This work was supported by JSPS KAKENHI Grant Numbers 17K05516
and 21K03466.
}

\appendix
\section{Spin--strain interaction with $C_{3v}$ symmetry}
When the strain tensors $\varepsilon_{ij}$ ($i,j = x, y ,z$) in the $C_{3v}$ reference frame are used
instead of those in the cubic ($XYZ$) reference frame, the strain-coupling coefficients
$A_{k,\varepsilon}$ in Eq.~(\ref{eqn:Hep}) are written as
\cite{Udvarhelyi18}
\begin{align}
& A_{u, \varepsilon} = \frac{1}{\sqrt{3}} [ h_{41} ( \varepsilon_{xx} + \varepsilon_{yy} )
+ h_{43} \varepsilon_{zz} ],
\nonumber \\
& A_{v, \varepsilon} = - \frac{1}{2} \left[ h_{16} \varepsilon_{zx}
- \frac{1}{2} h_{15} ( \varepsilon_{xx} - \varepsilon_{yy} ) \right],
\nonumber \\
& A_{zx, \varepsilon} = \frac{1}{2} \left[ h_{26} \varepsilon_{zx}
- \frac{1}{2} h_{25} ( \varepsilon_{xx} - \varepsilon_{yy} ) \right].
\label{eqn:Akep}
\end{align}
The three symmetry strain components $2 \varepsilon_{zz} - \varepsilon_{xx} - \varepsilon_{yy}$,
$\varepsilon_{xx} - \varepsilon_{yy}$, and $\varepsilon_{zx}$ are shown in Fig.~\ref{fig:1}(a) on
the right.
For the coupling-strength parameters $h_{\alpha \beta}$ here, the subscripts $\alpha \beta$
follow those in the complete form of the spin--strain interaction Hamiltonian in Ref.~5.
As shown in Fig.~\ref{fig:3}(a), $A_{v,\varepsilon}$ and $A_{zx,\varepsilon}$ are for the DQ and
SQ transitions, respectively.
The strain tensors are transformed to those in the cubic coordinate in Eq.~(\ref{eqn:Ak-ep}) as
\cite{Koga20b}
\begin{align}
& \varepsilon_{xx} + \varepsilon_{yy} = - \frac{2}{\sqrt{3}} \varepsilon_1
+ \frac{2}{3} \varepsilon_{\rm B},
\nonumber \\
& \varepsilon_{zz} = \frac{2}{\sqrt{3}} \varepsilon_1 + \frac{1}{3} \varepsilon_{\rm B},
\nonumber \\
& \varepsilon_{xx} - \varepsilon_{yy} = \frac{1}{\sqrt{3}} \varepsilon_{U_1}
+ \frac{2}{\sqrt{3}} \varepsilon_{U_2},
\nonumber \\
& \varepsilon_{zx} = \frac{1}{\sqrt{6}} \varepsilon_{U_1} - \frac{1}{\sqrt{6}} \varepsilon_{U_2},
\label{eqn:epij}
\end{align}
where $\varepsilon_B$ ($= \varepsilon_{xx} + \varepsilon_{yy} + \varepsilon_{zz}
= \varepsilon_{XX} + \varepsilon_{YY} + \varepsilon_{ZZ}$) represents a bulk strain.
From these equations, the five coupling-strength parameters in Eq.~(\ref{eqn:Ak-ep}) are related to
$h_{\alpha \beta}$ as
\begin{align}
& g_a = \frac{2}{3} ( - h_{41} + h_{43} ),~~g_b = \frac{ h_{15} - \sqrt{2} h_{16} }{4},
\nonumber \\
& g_c = \frac{ 2 h_{15} + \sqrt{2} h_{16} }{4},~~g_d = \frac{ - h_{25} + \sqrt{2} h_{26} }{ 4\sqrt{2} },
\nonumber \\
& g_e = \frac{ 2 h_{25} + \sqrt{2} h_{26} }{ 2\sqrt{2} }.
\label{eqn:gh}
\end{align}
On the other hand, the symmetry-allowed form of the spin--stress interaction Hamiltonian is also
given as $H_\sigma = \sum_k A_{k, \sigma} O_k$ ($k = u, v, zx$), which corresponds to
Eq.~(\ref{eqn:Hep}).
The stress-dependent coupling coefficients $A_{k, \sigma}$ are analogous to
$A_{k, \varepsilon}$ in Eq.~(\ref{eqn:Akep}).
According to Ref.~5, the spin--strain coupling-strength parameters $h_{\alpha \beta}$ are simply
replaced by the spin--stress ones $g_{\alpha \beta}$ as
\begin{align}
& A_{u, \sigma} = \frac{1}{\sqrt{3}} [ g_{41} ( \sigma_{xx} + \sigma_{yy} )
+ g_{43} \sigma_{zz} ],
\nonumber \\
& A_{v, \sigma} = - \frac{1}{2} \left[ g_{16} \sigma_{zx}
- \frac{1}{2} g_{15} ( \sigma_{xx} - \sigma_{yy} ) \right],
\nonumber \\
& A_{zx, \sigma} = \frac{1}{2} \left[ g_{26} \varepsilon_{zx}
- \frac{1}{2} g_{25} ( \sigma_{xx} - \sigma_{yy} ) \right].
\label{eqn:Aksig}
\end{align}
The stress tensors $\sigma_{ij}$ ($i,j = x,y,z$) can be transformed to those in the cubic ($XYZ$)
frame as well as the strain tensors in Eq.~(\ref{eqn:epij}).
Following the representations introduced in Ref.~5, the stress-dependent coupling coefficients
are rewritten as
\begin{align}
& A_{u, \sigma} = 2 a_2 \sigma_1,~~
A_{v, \sigma} = - \sqrt{3} ( b \sigma_{U_1} + c \sigma_{U_2} ),
\nonumber \\
& A_{zx, \sigma} = \sqrt{3} ( d \sigma_{U_1} + e \sigma_{U_2} ),
\end{align}
where the stress-field components are given by
\begin{align}
& \sigma_1 = ( \sigma_{YZ} + \sigma_{ZX} + \sigma_{XY} ) / \sqrt{3},
\nonumber \\
& \sigma_{U_1} = ( 2 \sigma_{ZZ} - \sigma_{XX} - \sigma_{YY} ) / \sqrt{3},
\nonumber \\
& \sigma_{U_2} = ( 2 \sigma_{XY} - \sigma_{YZ} - \sigma_{ZX} ) / \sqrt{3}.
\end{align}
The spin--stress coupling-strength parameters $\{ a_2, b, c, d, e \}$ are related to
$g_{\alpha \beta}$ in Eq.~(\ref{eqn:Aksig}) as
\cite{note1}
\begin{align}
& a_2 = \frac{ - g_{41} + g_{43} }{3},~~b= \frac{ - g_{15} + \sqrt{2} g_{16} }{12},
\nonumber \\
& c = \frac{ - 2 g_{15} - \sqrt{2} g_{16} }{12},~~d = \frac{ - g_{25} + \sqrt{2} g_{26} }{12},
\nonumber \\
& e = \frac{ - 2 g_{25} - \sqrt{2} g_{26} }{12}.
\label{eqn:spin-stress}
\end{align}
\par

Finally, we connect the coupling-strength parameters of the spin--strain interaction to those of
the spin--stress interaction using the elastic stiffness tensors $\{ C_{11}, C_{12} , C_{44} \}$ in the
cubic frame as
\begin{align}
& g_a = 4 a_2 C_{44},~~g_b = - 3 b ( C_{11} - C_{12} ),~~g_c = -6 c C_{44},
\nonumber \\
& g_d = 3 d ( C_{11} - C_{12} ) / \sqrt{2},~~g_e = - 6 \sqrt{2} e C_{44}.
\end{align}
Here, $g_b$ and $g_c$ ($b$ and $c$) are related to the phonon-driven DQ transitions, and
$g_d$ and $g_e$ ($d$ and $e$) are related to the SQ transitions.
It is obvious that the ratios of the spin--strain coupling-strength parameters are directly connected
to those of the spin--stress ones as
\begin{align}
\frac{ g_d }{ g_b } = - \frac{d}{ \sqrt{2} b },~~\frac{ g_e }{ g_c } = \frac{ \sqrt{2} e }{c}.
\label{eqn:SQ-DQ}
\end{align}

\section{Quadrupole--strain couplings for $\theta = \pi / 2$}
Here, we show the case of $\theta = \pi / 2$ (the magnetic field is perpendicular to $[111]$), and
this analytic solution is useful for our analysis.
\cite{Koga20b}
In Eq.~(\ref{eqn:bHl}), the eigenenergies are given by
\begin{align}
\bar{E}_1 = \frac{1}{2} ( - 1 - \alpha ),~~\bar{E}_2 = 1,~~\bar{E}_3 = \frac{1}{2} ( - 1 + \alpha ),
\label{eqn:energy3}
\end{align}
where $\alpha = \sqrt{ 9 + 4 \bar{b}^2 }$.
For the $\bar{b}$-dependent coefficients in Eq.~(\ref{eqn:psimu}),
\begin{align}
& C_{11} = C_{-1,1} = \frac{1}{\sqrt{2}} \sin \chi,~~C_{01} = \cos \chi,
\nonumber \\
& C_{12} = - C_{-1, 2} = \frac{1}{\sqrt{2}},~~C_{02} = 0
\nonumber \\
& C_{13} = C_{-1,3} = \frac{1}{\sqrt{2}} \cos \chi,~~C_{03} = - \sin \chi,
\end{align}
and $\chi$ is related to $\alpha$ as
\begin{align}
\cos \chi = \sqrt{ \frac{1}{2} \left( 1 + \frac{3}{\alpha} \right) },~~
\sin \chi = \sqrt{ \frac{1}{2} \left( 1- \frac{3}{\alpha} \right) }.
\end{align}
For $A_{\mu \nu}$, we obtain
\begin{align}
& A_{11} = - \frac{1}{ 2 \sqrt{3} } ( 1 + 3 \cos 2 \chi ) A_u
+ \frac{1}{2} ( 1 - \cos 2 \chi ) \cos 2 \phi \cdot A_v
\label{eqn:a11} \\
& A_{22} = \frac{1}{ \sqrt{3} } A_u - \cos 2 \phi \cdot A_v
\label{eqn:a22} \\
& A_{33} = - \frac{1} { 2 \sqrt{3} } ( 1 - 3 \cos 2 \chi ) A_u
+ \frac{1}{2} ( 1 + \cos 2 \chi ) \cos 2 \phi \cdot A_v \\
& A_{12} = - i \sin \chi \sin 2 \phi \cdot A_v + \cos \chi \cos \phi \cdot A_{zx} \\
& A_{13} = \frac{ \sqrt{3} }{2} \sin 2 \chi \cdot A_u + \frac{1}{2} \sin 2 \chi \cos 2 \phi \cdot A_v
- i \sin \phi \cdot A_{zx} \\
& A_{23} = i \cos \chi \sin 2 \phi \cdot A_v - \sin \chi \cos \phi \cdot A_{zx}.
\end{align}
When $\theta$ decreases from $\pi / 2$, $| C_{1 \mu} |$ and $| C_{-1 \mu} |$ become unequal
and $| C_{02} | \ne 0$.
As a consequence, $A_{zx}$ for the coupling with the $zx$ quadrupole also participates in the
diagonal matrix elements $A_{\mu \mu}$, and $A_{12}$ for the transition
$| \psi_1 \rangle \leftrightarrow | \psi_2 \rangle$ does not vanish at $\phi = \pi / 2$.
\par

In the weak coupling limit ($A_k / \omega \ll 1$, where $k = u, v,zx$) for $\theta = \pi / 2$,
Eq.~(\ref{eqn:a-k}) is simplified as
\begin{align}
\cos 2 \phi_B = \frac{ A_u }{ \sqrt{3} A_v } \frac{ 1 + \cos 2 \chi }{ 1 - ( \cos 2 \chi ) / 3 }~~
( A_L \rightarrow 0 ),
\end{align}
because $A_{zx}$ does not appear in Eq.~(\ref{eqn:a11}) or (\ref{eqn:a22}), namely, in the
longitudinal coupling $A_L$.
For $| A_u | \ll |A_v |$, we find that $\bar{P} (\varepsilon_{12} = 2 \omega)$ shows a
significant decrease as $\phi_B$ approaches $\pi / 4$, regardless of $| A_{zx} |$.
When $\theta$ is changed from $\pi / 2$, a finite $A_{zx}$ term in $A_L$ affects the $\phi$
dependence of $\bar{P}$.
As indicated in Fig.~\ref{fig:4}, the value of $\phi_B$ for $A_L \rightarrow 0$ shifts toward $\pi / 2$
with an increase in $a_{zx}$ ($= A_{zx} / A_v$).

\section{Reconsideration of $A_L \rightarrow 0$ for multiple components of strain fields}
Here, we discuss a necessary condition for the vanishment of the longitudinal quadrupole--strain
coupling $A_L \rightarrow 0$ used for our analysis of two-phonon transition probabilities.
In Eq.~(\ref{eqn:Ht}), we do not consider the relative phase shifts between the three
components of time-dependent oscillating strain fields.
This assumption holds for plane acoustic waves or surface acoustic waves with small decay.
Indeed, $A_L$ does not approach zero if there is a phase difference between two strain
components.
\par

Let us consider a surface acoustic wave propagating along the $[110]$ direction in the $XY$
plane and oscillating along the $Z$ direction with decay.
Here, the displacement of the transverse wave with wave number $k$ is simply expressed as
\begin{align}
u_Z = U e^{- q Z} \sin ( k \xi - \omega t ),~~\left( \xi = \frac{X + Y}{\sqrt{2}} , Z > 0 \right),
\label{eqn:SAW}
\end{align}
where $U$ is a real amplitude and $1 / q$ represents a penetration depth measured from the
surface at $Z = 0$.
Assuming that $q \ll k$, we neglect the longitudinal displacement component in the $[110]$
direction.
Accordingly, the following strain components are induced:
\begin{align}
& \varepsilon_{ZZ} = - q U e^{-q Z} \sin ( k \xi - \omega t ), \\
& \varepsilon_{YZ} = \varepsilon_{ZX} = \frac{ k U }{ 2 \sqrt{2} } e^{-q Z} \cos ( k \xi - \omega t).
\end{align}
When we represent the strain components as $\varepsilon_1 = a_1 \cos (k \xi - \omega t)$,
$\varepsilon_{U_1} = a_{U_1}\sin (k \xi - \omega t)$,
and $\varepsilon_{U_2} = a_{U_2} \cos (k \xi - \omega t)$ in Eq.~(\ref{eqn:Ak-ep}), we have
\begin{align}
a_1 = - a_{U_2} = \frac{k U}{\sqrt{6}},~~a_{U_1} = - \frac{2 q U}{\sqrt{3}}.
\end{align}
Using them in Eq.~(\ref{eqn:A-L}), we can obtain the time-dependent longitudinal
quadrupole--strain coupling written as
\begin{align}
A_L (t) = a_1 g_c ( \phi ) \cos \omega t + a_{U_1} g_s (\phi) \sin \omega t,
\end{align}
where
\begin{align}
& g_c ( \phi ) = \alpha_-^{(u)} g_a - \frac{ \alpha_-^{(v+)} g_c }{ \sqrt{3} } \cos 2 \phi
+ \frac{ \alpha_-^{(zx+)} g_e }{ \sqrt{6} } \cos \phi, \\
& g_s ( \phi ) = \frac{ \alpha_-^{(v+)} g_b }{ \sqrt{3} } \cos 2 \phi
+ \frac{ 2 \alpha_-^{(zx+)} g_d }{ \sqrt{6} } \cos \phi.
\end{align}
These equations lead to the amplitude of $A_L (t)$ as
\begin{align}
& | A_L | = \sqrt{ [ a_1 g_c ( \phi ) ]^2 + [ a_{U_1} g_s ( \phi ) ]^2 }
\nonumber \\
&~~~~~~
\propto k U \sqrt{ [ g_c ( \phi ) ]^2 + 8 \left( \frac{q}{k} \right)^2 [ g_s ( \phi ) ]^2 }.
\end{align}
When the penetration factor $q / k$ is sufficiently small, we can find the field-rotation angle
$\phi_B$ for $A_L \rightarrow 0$ or the minimum of $| A_L |$ as a function of $\phi$. 
If $q$ is comparable to $k$ in magnitude, it is necessary to generalize the model Hamiltonian in
Eq.~(\ref{eqn:Ht}) as
\begin{align}
H_{\rm eff} (t) \rightarrow \frac{1}{2}
\left(
\begin{array}{cc}
- \varepsilon_{12} + A_1 \cos ( \omega t + \theta_1 ) & A_T \cos \omega t \\
A_T^* \cos \omega t & \varepsilon_{12} + A_2 \cos ( \omega t + \theta_2 )
\end{array}
\right),
\end{align}
where the phase shifts of the oscillating fields are different between the two levels
($\theta_1 \ne \theta_2$).
It is straightforward to apply our present formulation to the extended model, which remains as a
future task.

\section{Comparison of two-level system with three-level system}
\begin{figure}
\begin{center}
\includegraphics[width=7cm,clip]{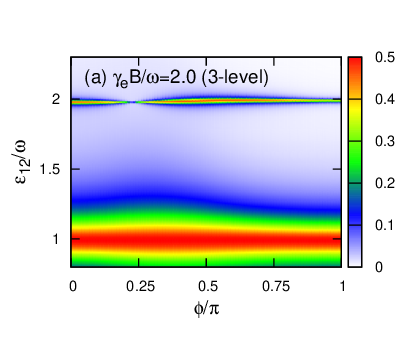}
\includegraphics[width=6cm,clip]{fig6b.eps}
\includegraphics[width=6cm,clip]{fig6c.eps}
\end{center}
\caption{
(Color online)
(a) Contour map of $\bar{P}_{ \psi_1 \rightarrow \psi_2 }$ as a function of $\phi$ and
$\varepsilon_{12} / \omega$ for the three-level system including finite $A_{13}$ and $A_{23}$ for
the quadrupole--strain couplings with $| \psi_3 \rangle$.
The parameters $D_0$, $B$, and $(A_u, A_v, A_{zx})$ are the same as given in Fig.~\ref{fig:5}(a).
(b) $\bar{P}_{ \psi_1 \rightarrow \psi_2 } (\phi)$ at $\varepsilon_{12} = 2 \omega$ for the two-level
($A_{13} = A_{23} = 0$) and three-level systems.
(c) Plots of the absolute values of various quadrupole--strain (QS) couplings  as a function of
$\phi$, normalized by $| A_v |$.
Here, $| A_L | = | A_{22} - A_{11} |$.
}
\label{fig:6}
\end{figure}
We check the robustness of the two-level treatment for the $S =1$ triplet considering the
contribution from $| \psi_3 \rangle$ in the highest sublevel.
Using Eq.~(\ref{eqn:H3}) and the corresponding Floquet Hamiltonian, this is demonstrated by
comparing $\bar{P}_{\psi_1 \rightarrow \psi_2}$ between the two-level system with
$A_{13} = A_{23} = 0$ and the three-level system including finite $A_{13}$ and $A_{23}$.
Typical numerical results are given as the contour maps of $\bar{P} (\phi, \varepsilon_{12})$
in Fig.~\ref{fig:5}(a) for the former and those in Fig.~\ref{fig:6}(a) for the latter.
From the comparison of the two panels, it is evident that the overviews of
$\bar{P}(\phi, \varepsilon_{12} / \omega)$ are almost identical.
In particular, $\bar{P} (\phi)$ exhibits a node in the horizontal line at
$\varepsilon_{12} / \omega \simeq 2$ at almost the same position.
This result confirms that the phonon-mediated transition probability for
$| \psi_1 \rangle \rightarrow | \psi_2 \rangle$ in the spin triplet can be well described by the
two-level Hamiltonian.
\par

For a more precise comparison, the data of $\bar{P} (\varepsilon_{12} / \omega = 2)$ are plotted
as a function of $\phi$ for $\gamma_e B / \omega = 2.0$ in Fig.~\ref{fig:6}(b).
In $0 < \phi < \pi / 2$, $\bar{P} (\phi)$ is almost identical in the two- and three-level systems.
The contribution from $| \psi_3 \rangle$ is negligible owing to $| A_{12} | > | A_{13} |$ for
smaller values of $\phi$, as shown in Fig.~\ref{fig:6}(c).
On the other hand, the increase in $| A_{13} |$ for $\phi / \pi \gtrsim 0.5$ causes large deviations
between the two plots of $\bar{P} (\phi)$ in Fig.~\ref{fig:6}(b).
We also find that $\bar{P} (\phi)$ is not greatly dependent on $| A_{23} |$, indicating that the
quadrupole--strain coupling between $| \psi_2 \rangle$ and
$| \psi_3 \rangle$ is not relevant in the $| \psi_1 \rangle \rightarrow | \psi_2 \rangle$ transition.
Thus, the transition probability can be well described within the two lowest-lying states of the
$S=1$ triplet, and the argument on the two-level system in Sect.~4.2 is useful for evaluating the
quadrupole--strain couplings associated with the field-angle-resolved MAR.



\begin{thebibliography}{99}

\bibitem{VanVleck40}
  J. H. Van Vleck,
  Phys. Rev. {\bf 57}, 426 (1940).

\bibitem{Donoho64}
  P. L. Donoho,
  Phys. Rev. {\bf 133}, A1080 (1964).
  
\bibitem{Barfuss15}
  A. Barfuss, J. Teissier, E. Neu, A. Nunnenkamp, and P. Maletinsky,
  Nat. Phys. {\bf 11}, 820 (2015).

\bibitem{Lee17}
  D. Lee, K. W. Lee, J. V. Cady, P. Ovartchaiyapong, and A. C. Bleszynski Jayich,
  J. Opt. {\bf 19}, 033001 (2017).

\bibitem{Udvarhelyi18}
  P. Udvarhelyi, V. O. Shkolnikov, A. Gali, G. Burkard, and A. P$\acute{\rm a}$lyi,
  Phys. Rev. B {\bf 98}, 075201 (2018).
  
\bibitem{Barfuss19}
  A. Barfuss, M. Kasperczyk, J. K\"{o}lbl, and P. Maletinsky,
  Phys. Rev. B {\bf 99}, 174102 (2019).

\bibitem{Chen20}
  H. Y. Chen, S. A. Bhave, and G. D. Fuchs,
  Phys. Rev. Appl. {\bf 13}, 054068 (2020).

\bibitem{Mitsumoto14}
  K. Mitsumoto, M. Akatsu, S. Baba, R. Takasu, Y. Nemoto, T. Goto, H. Yamada-Kaneta,
  Y. Furumura, H. Saito, K. Kashima, and Y. Saito,
  J. Phys. Soc. Jpn. {\bf 83}, 034702 (2014).

\bibitem{Kiel72}
  A. Kiel and W. B. Mims,
  Phys. Rev. B {\bf 5}, 803 (1972).
  
\bibitem{Mims76}
  W. B. Mims,
  {\it The Linear Electric Field Effect in Paramagnetic Resonance}
  (Oxford University Press, Oxford, U. K., 1976).
  
\bibitem{VanOort90}
  E. Van Oort and M. Glasbeek,
  Chem. Phys. Lett. {\bf 168}, 529 (1990).

\bibitem{Doherty12}
  M. W. Doherty, F. Dolde, H. Fedder, F. Jelezko, J. Wrachtrup, N. B. Manson, and
  L. C. L. Hollenberg,
  Phys. Rev. B {\bf 85}, 205203 (2012).
  
\bibitem{Matsumoto17}
  M. Matsumoto, K. Chimata, and M. Koga,
  J. Phys. Soc. Jpn. {\bf 86}, 034704 (2017).
  
\bibitem{MacQuarrie13}
  E. R. MacQuarrie, T. A. Gosavi, N. R. Jungwirth, S. A. Bhave, and G. D. Fuchs,
  Phys. Rev. Lett. {\bf 111}, 227602 (2013).
  
\bibitem{Klimov14}
  P. V. Klimov, A. L. Falk, B. B. Buckley, and D. D. Awschalom,
  Phys. Rev. Lett. {\bf 112}, 087601 (2014).
  
\bibitem{Ovartchaiyapong14}
  P. Ovartchaiyapong, K. W. Lee, B. A. Myers, and A. C. Bleszynski Jayich,
  Nat. Commun. {\bf 5}, 4429 (2014).
   
\bibitem{Kepesidis13}
  K. V. Kepesidis, S. D. Bennett, S. Portolan, M. D. Lukin, and P. Rabl,
  Phys. Rev. B {\bf 88}, 064105 (2013).
  
\bibitem{Golter16}
  D. A. Golter, T. Oo, M. Amezcua, K. A. Stewart, and H. Wang,
  Phys. Rev. Lett. {\bf 116}, 143602 (2016).

\bibitem{Chen18}
  H. Y. Chen, E. R. MacQuarrie, and G. D. Fuchs,
  Phys. Rev. Lett. {\bf 120}, 167401 (2018).
    
\bibitem{Balasubramanian09}
  G. Balasubramanian, P. Neumann, D. Twitchen, M. Markham, R. Kolesov, N. Mizuochi, J. Isoya,
  J. Achard, J. Beck, J. Tissler, V. Jacques, P. R. Hemmer, F. Jelezko, and J. Wrachtrup,
  Nat. Mater. {\bf 8}, 383 (2009).
  
\bibitem{Mizuochi09}
  N. Mizuochi, P. Neumann, F. Rempp, J. Beck, V. Jacques, P. Siyushev, K. Nakamura,
  D. J. Twitchen, H. Watanabe, S. Yamasaki, F. Jelezko, and J. Wrachtrup,
  Phys. Rev. B {\bf 80}, 041201(R) (2009).

\bibitem{Herbschleb19}
  E. D. Herbschleb, H. Kato, Y. Maruyama, T. Danjo, T. Makino, S. Yamasaki, I. Ohki, K. Hayashi,
  H. Morishita, M. Fujiwara, and N. Mizuochi,
  Nat. Commun. {\bf 10}, 3766 (2019).

\bibitem{Suter17}
  D. Suter and F. Jelezko,
  Prog. Nucl. Magn. Reson. Spectrosc. {\bf 98-99}, 50 (2017).
  
\bibitem{Degen17}
  C. L. Degen, F. Reinhard, and P. Cappellaro,
  Rev. Mod. Phys. {\bf 89}, 035002 (2017).
  
\bibitem{Rembold20}
  P. Rembold, N. Oshnik, M. M. M\"{u}ller, S. Montangero, T. Calarco, and E. Neu,
  AVS Quantum Sci. {\bf 2}, 024701 (2020).
  
\bibitem{Ruf21}
  M. Ruf, N. H. Wan, H. Choi, D. Englund, and R. Hanson,
  J. Appl. Phys. {\bf 130}, 070901 (2021).
  
\bibitem{MacQuarrie15}
  E. R. MacQuarrie, T. A. Gosavi, A. M. Moehle, N. R. Jungwirth, S. A. Bhave, and G. D. Fucks,
  Optica {\bf 2}, 233 (2015).
  
\bibitem{Mamin14}
  H. J. Mamin, M. H. Sherwood, M. Kim, C. T. Rettner, K. Ohno, D. D. Awschalom, and D. Rugar,
  Phys. Rev. Lett. {\bf 113}, 030803 (2014).
  
\bibitem{Bauch18}
  E. Bauch, C. A. Hart, J. M. Schloss, M. J. Turner, J. F. Barry, P. Kehayias, S. Singh, and
  R. L. Walsworth,
  Phys. Rev. X {\bf 8}, 031025 (2018).
  
\bibitem{Barry20}
  J. F. Barry, J. M. Schloss, E. Bauch, M. J. Turner, C. A. Hart, L. M. Pham, and R. L. Walsworth,
  Rev. Mod. Phys. {\bf 92}, 015004 (2020).
  
\bibitem{Yamaguchi20}
  T. Yamaguchi, Y. Matsuzaki, S. Saijo, H. Watanabe, N. Mizuochi, and J. Ishi-Hayase,
  Jpn. J. Appl. Phys. {\bf 59}, 110907 (2020).
  
\bibitem{Dong19}
  X.-L. Dong and P.-B. Li,
  Phys. Rev. A {\bf 100}, 043825 (2019).

\bibitem{Michl19}
  J. Michl, J. Steiner, A. Denisenko, A. B\"{u}lau, A Zimmermann, K. Nakamura,
  H. Sumiya, S. Onoda, P. Neumann, J. Isoya, and J. Wrachtrup,
  Nano Lett. {\bf 19}, 4904 (2019).

\bibitem{Kehayias19}
  P. Kehayias, M. J. Turner, R. Trubko, J. M. Schloss, C. A. Hart, M. Wesson, D. R. Glenn, and
  R. L. Walsworth,
  Phys. Rev. B {\bf 100}, 174103 (2019).

\bibitem{Koga20b}
  M. Koga and M. Matsumoto,
  J. Phys. Soc. Jpn. {\bf 89}, 113701 (2020).
  
\bibitem{Gromov00}
  I. Gromov and A. Schweiger, J. Magn. Reson. {\bf 146}, 110 (2000).
  
\bibitem{Koga20a}
  M. Koga and M. Matsumoto,
  J. Phys. Soc. Jpn. {\bf 89}, 024701 (2020).
  
\bibitem{Matsumoto20}
  M. Matsumoto and M. Koga,
  J. Phys. Soc. Jpn. {\bf 89}, 084702 (2020).
  
\bibitem{Valimaa21}
  A. V\"{a}limaa, W. Crump, M. Kervinen, and M. A. Sillanp\"{a}\"{a},
  Phys. Rev. Appl. {\bf 17}, 064003 (2022).
  
\bibitem{Shirley65}
  J. H. Shirley,
  Phys. Rev. {\bf 138}, B979 (1965).
  
\bibitem{Son09}
  S.-K. Son, S. Han, and S.-I Chu,
  Phys. Rev. A {\bf 79}, 032301 (2009).
    
\bibitem{Tetienne12}
  J.-P. Tetienne, L. Rondin, P. Spinicelli, M. Chipaux, T. Debuisschert, J.-F. Roch, and V. Jacques,
  New J. Phys. {\bf 14}, 103033 (2012).

\bibitem{Michl14}
  J. Michl, T. Teraji, S. Zaiser, I. Jakobi, G. Waldherr, F. Dolde, P. Neumann, M. W. Doherty,
  N. B. Manson, J. Isoya, and J. Wrachtrup,
  Appl. Phys. Lett. {\bf 104}, 102407 (2014).

\bibitem{Lesik14}
  M. Lesik, J.-P. Tetienne, A. Tallaire, J. Achard, V. Mille, A. Gicquel, J.-F. Roch, and V. Jacques,
  Appl. Phys. Lett. {\bf 104}, 113107 (2014).

\bibitem{Miyazaki14}
  T. Miyazaki, Y. Miyamoto, T. Makino, H. Kato, S. Yamasaki, T. Fukui, Y. Doi, N. Tokuda,
  M. Hatano, and N. Mizuochi,
  Appl. Phys. Lett. {\bf 105}, 261601 (2014).

\bibitem{Ozawa17}
  H. Ozawa, K. Tahara, H. Ishiwata, M. Hatano, and T. Iwasaki,
  Appl. Phys. Express {\bf 10}, 045501 (2017).

\bibitem{Mizuno21}
  K. Mizuno, M. Nakajima, H. Ishiwata, M. Hatano, and T. Iwasaki,
  Appl. Phys. Express {\bf 14}, 032001 (2021).
  
\bibitem{Tatsuishi21}
  T. Tatsuishi, K. Kanehisa, T. Kageura, T. Sonoda, Y. Hata, K. Kawakatsu, T. Tanii, S. Onoda,
  A. Stacey, S. Kono and H. Kawarada,
  Carbon {\bf 180}, 127 (2021).
  
\bibitem{Sohn18}
  Y.-I. Sohn, S. Meesala, B. Pingault, H. A. Atikian, J. Holzgrafe, M. G\"{u}ndo\v{g}an, C. Stavrakas,
  M. J. Stanley, A. Sipahigil, J. Choi, M. Zhang, J. L. Pacheco, J. Abraham, E. Bielejec, M. D. Lukin,
  M. Atat\"{u}re, and M. Lon\v{c}ar,
  Nat. Commun. {\bf 9}, 2012 (2018).
  
\bibitem{Meesala18}
  S. Meesala, Y.-I. Sohn, B. Pingault, L. Shao, H. A. Atikian, J. Holzgrafe, M. G\"{u}ndo\v{g}an,
  C. Stavrakas, A. Sipahigil, C. Chia, R. Evans, M. J. Burek, M. Zhang, L. Wu, J. L. Pacheco,
  J. Abraham, E. Bielejec, M. D. Lukin, M. Atat\"{u}re, and M. Lon\v{c}ar,
  Phys. Rev. B {\bf 97}, 205444 (2018).
 
\bibitem{Maity20}
  S. Maity, L. Shao, S. Bogdanovi\'{c}, S. Meesala, Y.-I. Sohn, N. Sinclair, B. Pingault,
  M. Chalupnik, C. Chia, L. Zheng, K. Lai, and M. Lon\v{c}ar,
  Nat. Commun. {\bf 11}, 193 (2020).
  
\bibitem{Whiteley19a}
  S. J. Whiteley, G. Wolfowicz, C. P. Anderson, A. Bourassa, H. Ma, M. Ye, G. Koolstra,
  K. J. Satzinger, M. V. Holt, F. J. Heremans, A. N. Cleland, D. I. Schuster, G. Galli, and
  D. D. Awschalom,
  Nat. Phys. {\bf 15}, 490 (2019).

\bibitem{Soltamov19}
  V. A. Soltamov, C. Kasper, A. V. Poshakinskiy, A. N. Anisimov, E. N. Mokhov, A. Sperlich,
  S. A. Tarasenko, P. G. Baranov, G. V. Astakhov, and V. Dyakonov,
  Nat. Commun. {\bf 10}, 1678 (2019).
  
\bibitem{Whiteley19b}
  S. J. Whiteley, F. J. Heremans, G. Wolfowicz, D. D. Awschalom, and M. V. Holt,
  Nat. Commun. {\bf 10}, 3386 (2019).
    
\bibitem{Hernandez-Minguez20}
  A. Hern\'{a}ndez-M\'{i}nguez, A. V. Poshakinskiy, M. Hollenbach, P. V. Santos, and
  G. V. Astakhov,
  Phys. Rev. Lett. {\bf 125}, 107702 (2020).
  
\bibitem{Singh22}
  H. Singh, M. A. Hollberg, A. N. Anisimov, P. G. Baranov, and D. Suter,
  Phys. Rev. Res. {\bf 4}, 023022 (2022).

\bibitem{note1}
  In Ref.~7, $b'$ and $c'$ are used instead of $d$ and $e$ for the spin--stress coupling-strength
  parameters, respectively.

\end{thebibliography}
\end{document}